# Flight Control System Design for Autonomous Aerial Surveys of Volcanoes

Zhe Shen (s172005)

Master of Science in Electrical Engineering

Automation and Robotics

2019

DTU Electrical Engineering

Department of Electrical Engineering

**Report written by:**

Zhe Shen (s172005)

**Advisor(s):**

Roberto Galeazzi, Associate Professor at Department of Electrical Engineering at the Technical University of Denmark, Denmark.

Takeshi Tsuchiya, Professor at Department of Aeronautics and Astronautics at the University of Tokyo, Japan.

**DTU Electrical Engineering**

Technical University of Denmark

2800 Kgs. Lyngby

Denmark

elektro@elektro.dtu.dk

Remarks: This report is submitted as partial fulfillment of the requirements for graduation in the above education at the Technical University of Denmark. Some control methods invented may be recollected and modified for formal journal publication.

# Summary

The controller for a quadrotor working in severe environment is developed in this study. Here, the severe environment indicates the temperature-varying air near the volcano. The controller overcomes the intensively changing temperature above the crater of the volcano which biases the nominal dynamics (25℃).

The target Volcano is picked as Satsuma-iojima located in a tiny insular South to the mainland of Japan. The temperature distribution is contributed from previous research by Geological Survey of Japan.

To guarantee that the control signal is under the input saturation, a path planning method is developed. Picking the eigenvalue for a system with a moving reference is a novel topic; the method to develop a controller with specific requirement is created for the first time. This method might be referred as a standard way for designers/engineers in developing controller with a moving target/reference in further study.

In controlling part, a state feedback controller is designed to stabilize the height of the quadrotor. The eigenvalue of the feedback controller is picked based on the method developed in Chapter ‘Path Planning’. And a PID controller is designed to control the attitude. The result of these are verified in a simulator written in MATLAB.

At last, a Kalman filter is applied in height control to combine the measurement noise from IMU and laser scanner and the system noise caused by the changing temperature. Another reason for developing a Kalman filter is that IMU readout provides with acceleration. While the velocity is not achieved by sensor directly.

The result of height control with a Kalman filter is verified in MATLAB Simulink.

# Acknowledgements

During World War II, Niels Bohr was forced to move back to his hometown, Copenhagen, from Germany, thus being separated from his outstanding German student and sincere friend, Werner Heisenberg. Despite high political sensitivity, Heisenberg visited his cherished supervisor, Bohr, when Denmark was occupied by Germany years later, in 1941.

This visit, however, was one of the most famous unsuccessful meetings in history. The friendship between them came to an end after that eventful day.

After 50 years of Bohr's death, some unsent letters to Heisenberg came to light for the first time. The real emotion hidden in this physicist was revealed by these Bohr's manuscript; Bohr chose to forgive Heisenberg with a sophisticated potential agreement. While the meeting in 1941 remains in a mystery even for today.

Learned from this, we people today are supposed to express our appreciation as sincere as we can, for as soon as we can.

Before leaving Denmark, I will say thank you to Roberto Galeazzi. He always provides advice in a way that I can understand. Without his help, this study would hardly start. Besides, I need to appreciate the colleagues in Takeshi Tsuchiya's lab. They helped a lot in simulator environment. The raw data of the Volcano is kindly provided by Nobuo Matsushima from Geological Survey of Japan.

Also, the care from friends and family can never be ignored. Thank you all for being with me!

# Contents

# CHAPTER 1

# The Volcano

The target Volcano where the quadrotor is supposed to be controlled on in this study is Satsuma-iojima Volcano. Satsuma-iojima Volcano (Fig.1.1) is located at an island, Satsuma-iojima insula, about 50 km to the south of Satsuma Peninsula.

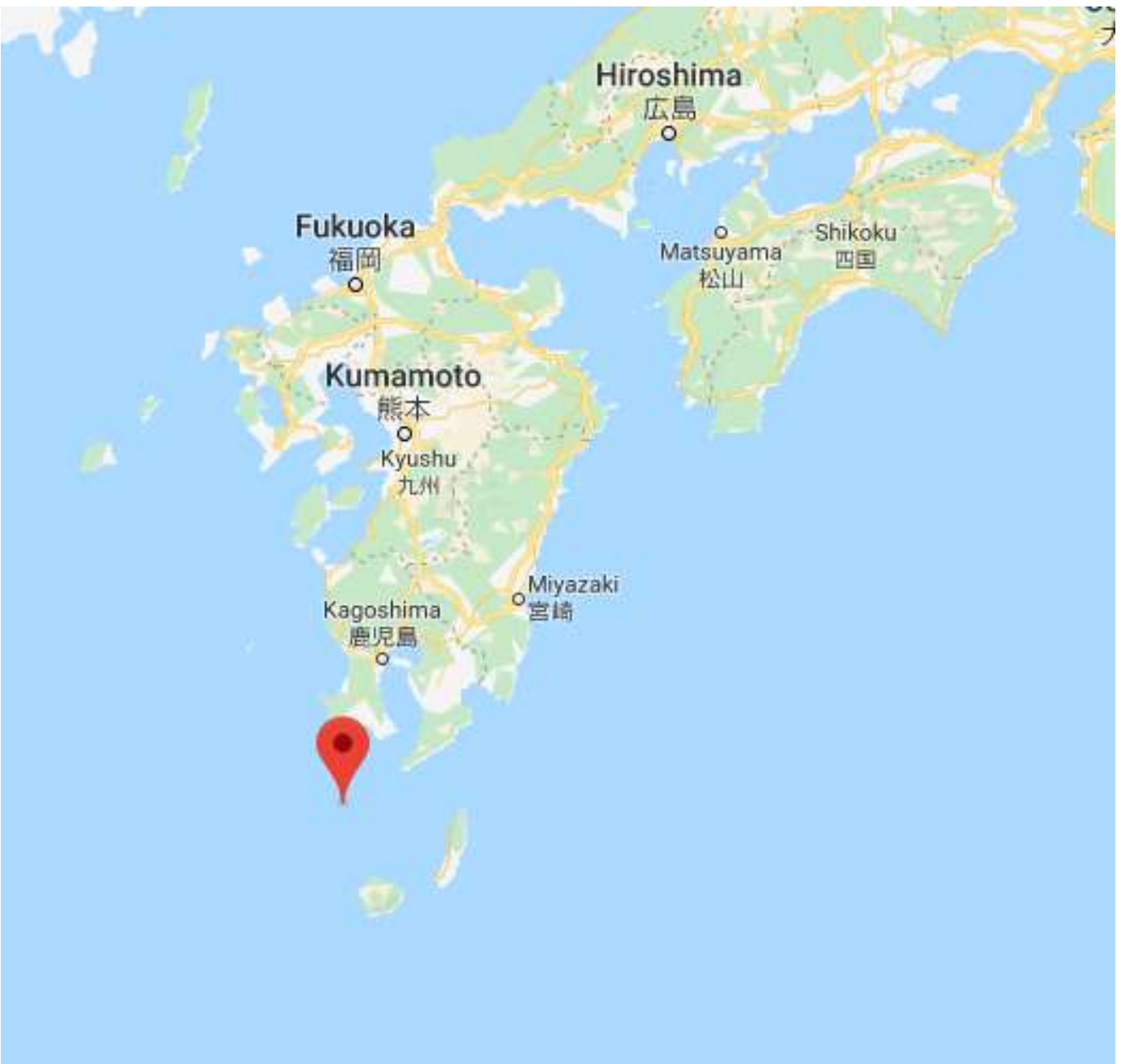

Figure 1.1. Satsuma-iojima insula, about 50 km to the south of Satsuma Peninsula. This map is provided by Google Map 2019.

The data of the temperature distribution of Satsuma-iojima has been collected by previous study. In Nobuo Matsushima's research (Estimation of permeability and degassing depth of Iwodake volcano at Satsuma-Iwojima, Japan, by thermal and numerical simulation, J. Volcanol. Geotherm. Res, 202, 2011, 167-177), it is presented in the form of a temperature map (Fig1.2).

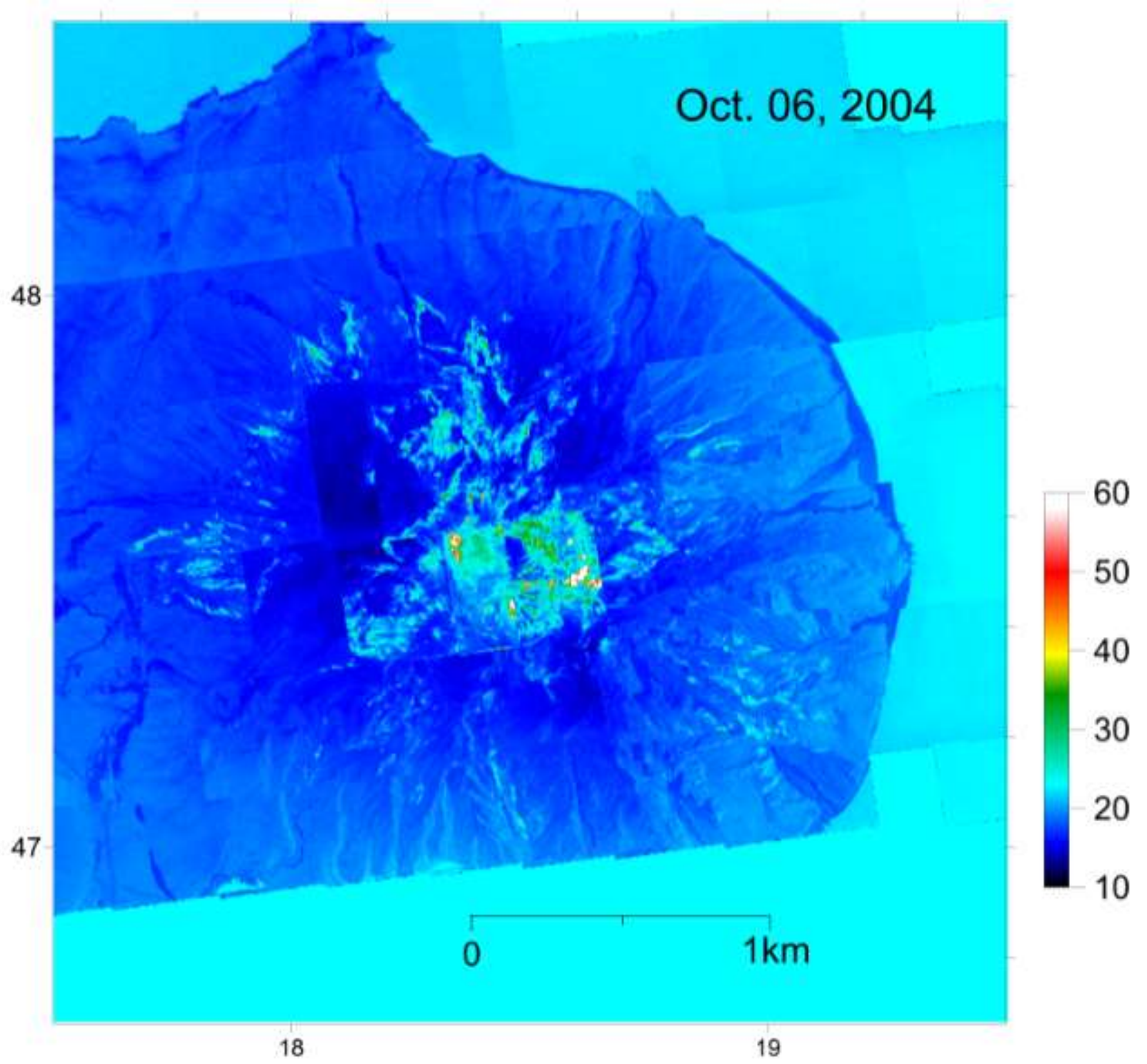

Figure 1.2. The temperature distribution of Satsuma-iojima insula. This is the temperature on the data of Oct. 06. 2004.
(https://gbank.gsj.jp/volcano/Act_Vol/satsumaioujima/vr/edoc/005.html)

The high temperature part in the map are the areas near the crater of Satsuma-iojima Volcano. Another data of temperature distribution in 1997 (Fig.1.3) is published on the website of Geological Survey of Japan.

It can be accessed from the temperature distribution in 1997 that the temperature near the crater varies from 185℃ to 885℃. And this information will be used later in this research.

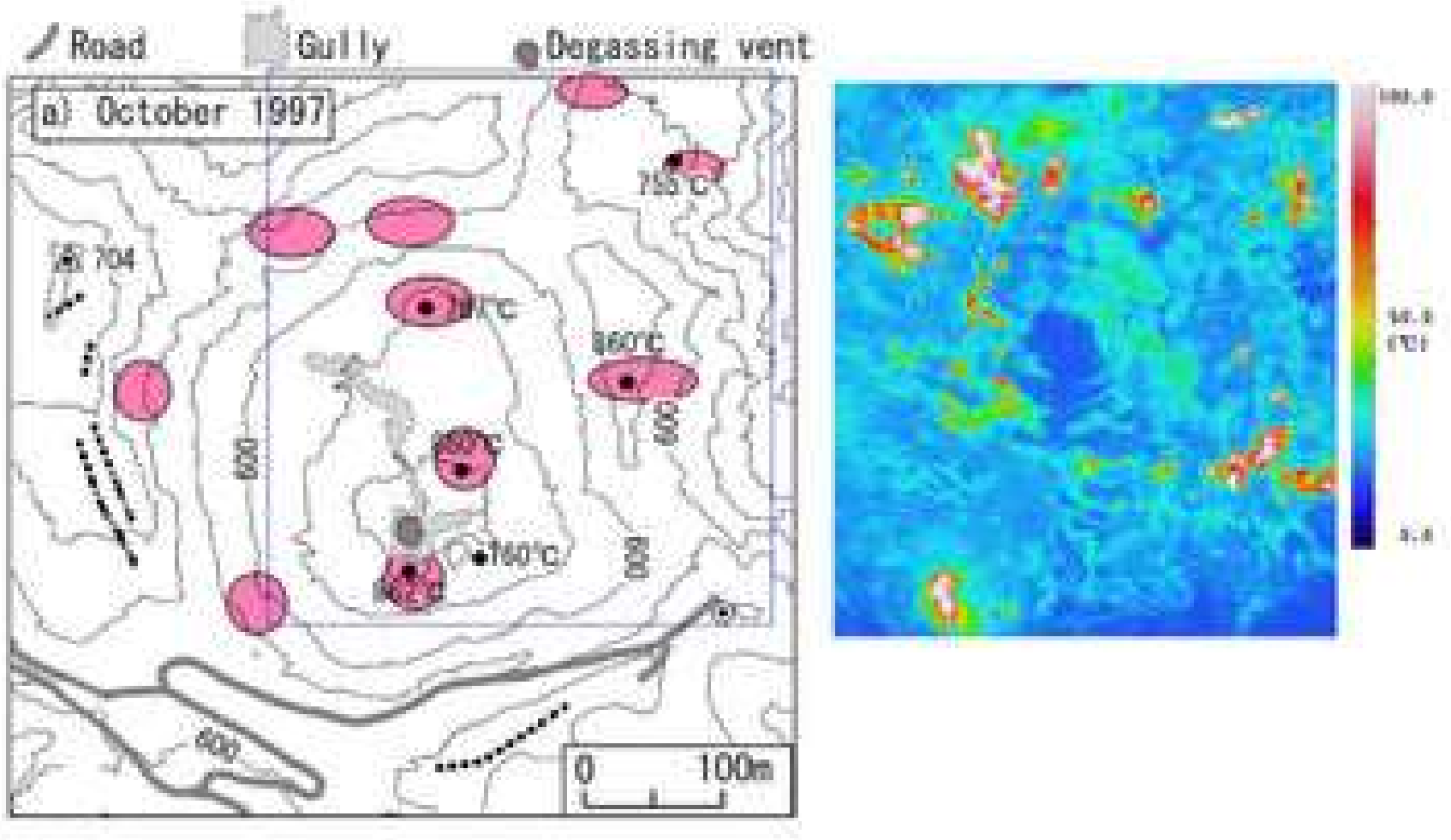

Figure 1.3. The temperature distribution of Satsuma-iojima insula in October 1997.

The well-controlled quadrotor is supposed to lift to a specific height before hovering at that height. And it should translate for a distance to a position over the crater before hovering again. The next process is to translate back before hovering back to that position. Finally, the quadrotor descends back to the ground (Fig.1.4).

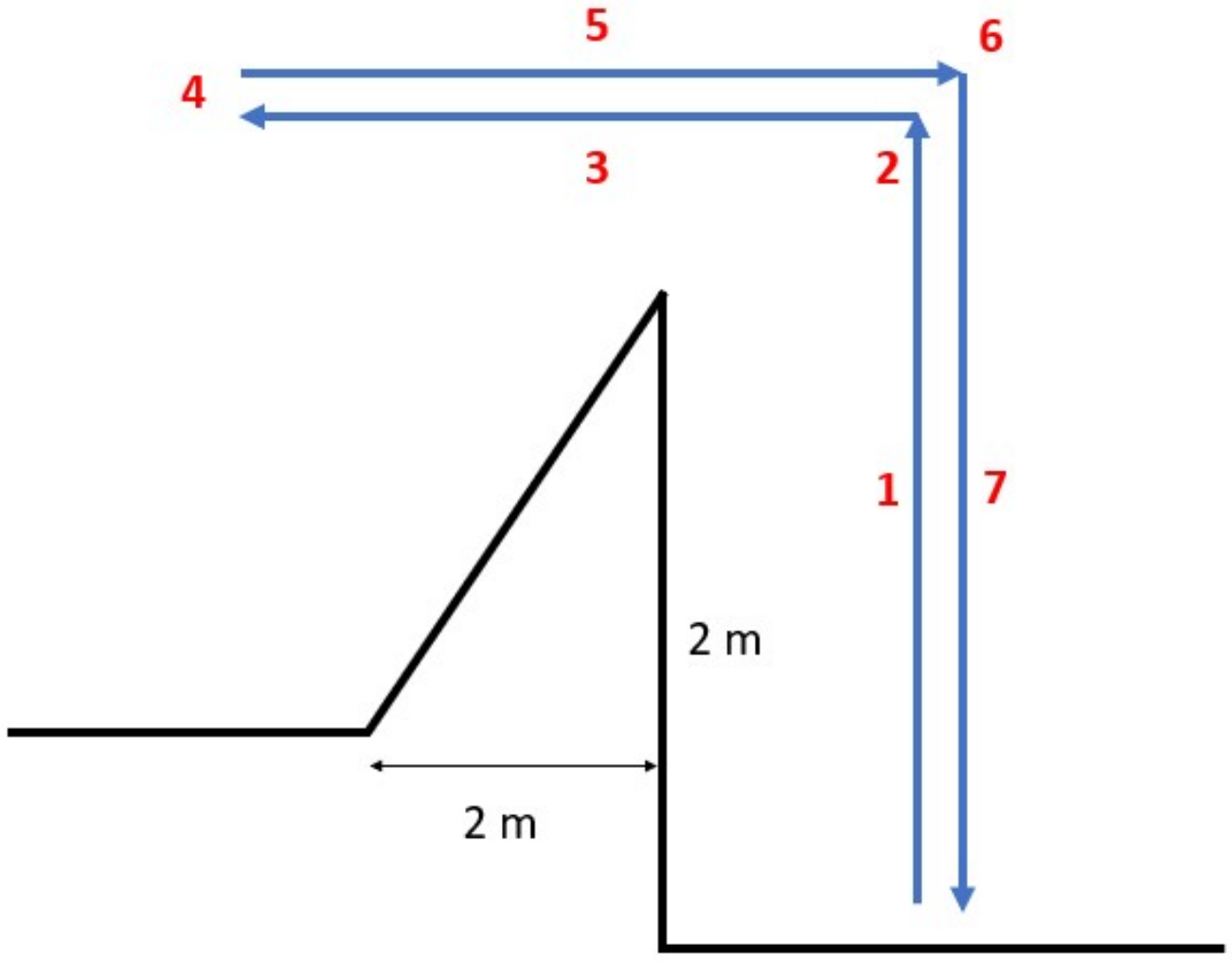

Figure 1.4. The movements which the quadrotor is supposed to accomplished.

The main difficulty for controlling the quadrotor here is the dynamic is uncertain caused by the intensively changing temperature varying from 185℃ to 885℃ near the crater of the Volcano.

CHAPTER 2

# Model the Quadrotor

Before making further steps in providing the dynamics of the quadrotor, the properties of the quadrotor are to be declared.

The mass of the quadrotor is:

$$m = 0.18 \text{ kg}$$

The length of each arm is:

$$L = 0.086 \text{ m}$$

The moment of inertia of the quadrotor is:

$$I = \begin{bmatrix} 0.00025 & 0 & 2.55 \times 10^{-6} \\ 0 & 0.000232 & 0 \\ 2.55 \times 10^{-6} & 0 & 0.0003738 \end{bmatrix} kg \cdot m^2$$

The gravitational acceleration is regarded as a constant:

$$g = 9.81 \text{ N/kg}$$

Since the angular velocity of the motor has an upper bound, the maximum of the thrust (input saturation) provided by each motor is:

$$F_{max} = 0.5\, mg$$

The thrust and the drag moment are proportional to the square of angular velocity of each motor:

$$F = k_F \cdot w^2$$

$$M = k_M \cdot w^2$$

The dynamics of the quadrotor will be discussed next where the changing temperature is not considered here.

Based on Newtown's Second Law,

$$\ddot{Z} = \frac{1}{m}(F_1 + F_2 + F_3 + F_4) - g \tag{2.1}$$

Here, Z is the altitude (height) of the quadrotor. Each of 4 F is the thrust provided by relevant motor. m is the mass of the quadrotor. g is the gravitational acceleration. The requirement for this formula is the attitude is well-controlled; the 4 thrusts are in the direction opposite the gravity force.

Note that in the simulator environment, there is no noise introduced. So that this formula is accurate. It fits the environment in Simulator and the knowledge in Chapter 'Basic Control'. While the dynamics equation in the environment near the volcano is remodeled in a different equation which we will discuss in Chapter 'Kalman Filter'.

We may rewrite it in the form of state space equation.

$$\begin{pmatrix} \dot{z} \\ \ddot{z} \end{pmatrix} = \begin{bmatrix} 0 & 1 \\ 0 & 0 \end{bmatrix} \begin{pmatrix} z \\ \dot{z} \end{pmatrix} + \begin{bmatrix} 0 & 0 & 0 & 0 \\ \frac{1}{m} & \frac{1}{m} & \frac{1}{m} & \frac{1}{m} \end{bmatrix} \begin{pmatrix} F_1 \\ F_2 \\ F_3 \\ F_4 \end{pmatrix} + \begin{pmatrix} 0 \\ -g \end{pmatrix}$$

or $$\dot{q} = \mathrm{A} \cdot \mathrm{q} + \mathrm{B} \cdot \mathrm{u} + \mathrm{v} \tag{2.2}$$

Also,

$$m\,\ddot{r} = \begin{pmatrix} 0 \\ 0 \\ -mg \end{pmatrix} + R \begin{pmatrix} 0 \\ 0 \\ F_1 + F_2 + F_3 + F_4 \end{pmatrix} \tag{2.3}$$

Here, r is a 3 by 1 position vector (x, y and z). R is the rotational vector, making the thrusts decompose into 3 parts in the directions of x, y and z.

Besides, based on Euler's equations (rigid body dynamics),

$$I \begin{pmatrix} \dot{p} \\ \dot{q} \\ \dot{r} \end{pmatrix} = \begin{bmatrix} L(F_2 - F_4) \\ L(F_3 - F_1) \\ M_1 - M_2 + M_3 - M_4 \end{bmatrix} - \begin{bmatrix} p \\ q \\ r \end{bmatrix} \times I \begin{bmatrix} p \\ q \\ r \end{bmatrix} \tag{2.4}$$

Here, I is the moment of inertia of the quadrotor. p, q and r are the attitude of the quadrotor (roll, pitch and yaw). L is the length of each arm. Each of M is the relevant drag moment provided by each motor.

Note that both M and F are proportional to the square of angular velocity. We may rewrite the Euler's equations (2.4),

$$I \begin{pmatrix} \dot{p} \\ \dot{q} \\ \dot{r} \end{pmatrix} = \begin{bmatrix} 0 & L & 0 & -L \\ -L & 0 & L & 0 \\ r & -r & r & -r \end{bmatrix} \begin{pmatrix} F_1 \\ F_2 \\ F_3 \\ F_4 \end{pmatrix} - \begin{bmatrix} p \\ q \\ r \end{bmatrix} \times I \begin{bmatrix} p \\ q \\ r \end{bmatrix} \tag{2.5}$$

Here, r is the ratio of M to F; this ratio equals to the ratio of $k_M$ to $k_F$.

And at 25℃,

$$k_M = 1.5 \times 10^{-9} N \cdot m/rpm^2 \tag{2.6}$$

$$k_F = 6.11 \times 10^{-8} N \cdot m/rpm^2 \tag{2.7}$$

$$\gamma = \frac{k_M}{k_F} = 0.0245 \tag{2.8}$$

# CHAPTER 3

# Path Planning

## Path Planning Algorithm

The purpose of developing path planning is to decrease the value of control signal.

As being mentioned in previous chapter 'Model the Quadrotor', the input has an upper bound value (input saturation). For the cases where the difference between target/reference and the initial position is relatively large, the calculated input needed for controlling the quadrotor may exceed this saturation value, making the real movement undesired in the designing purpose.

To decrease the difference in controlling, the reference would also move from initial position to final target rather than being fixed at final target. With a moving reference, the difference between quadrotor and the reference at each time point is less than the case where the reference is fixed at target.

We call the process in defining the way the reference moves path planning.

Overall, path planning witness 5 steps:

Step 1.

Link the initial position and the target position with a straight line. And calculate the length of the line (L).

Step 2.

Assume that the reference is moving with a constant velocity. Pick a proper average velocity of the reference (v). And calculate the total time for the reference moving from initial position to final target.

$$t_{total} = \frac{L}{v}$$

Step 3.

Assume that the moving function of the reference is a 7-order polynomial.

$$x(t) = a_0 + a_1 t + a_2 t^2 + a_3 t^3 + a_4 t^4 + a_5 t^5 + a_6 t^6 + a_7 t^7$$

Here, $a_0$ to $a_7$ are the coefficients.

Step 4.

Determine 8 coefficients, $a_0$ to $a_7$, based on the following 8 constraints:

a. $x(t = 0) = 0$
b. $x(t = t_{total}) = L$
c. velocity ($\dot{x}$), acceleration ($\ddot{x}$), jerk ($\dddot{x}$),…, $x^{(6)}$ are continuous at time $t_{total}$.

Step 5.

So, the coefficients are well-determined now. The reference is moving based on this polynomial.

$$x(t) = a_0 + a_1 t + a_2 t^2 + a_3 t^3 + a_4 t^4 + a_5 t^5 + a_6 t^6 + a_7 t^7$$

Both reasons for picking a 7-order polynomial in step 1 and for letting physical quantities continuous in step 4 are to make the reference move as smoothly as possible.

In my thesis, the average velocity is picked as 0.5 m/s. The generated path and generated velocity are plotted in blue lines. (Fig.3.1 and Fig.3.2)

It can be clearly seen that the blue lines are smooth even at the turning points since our algorithm doesn’t allow a sudden change in trajectory or velocity planned.

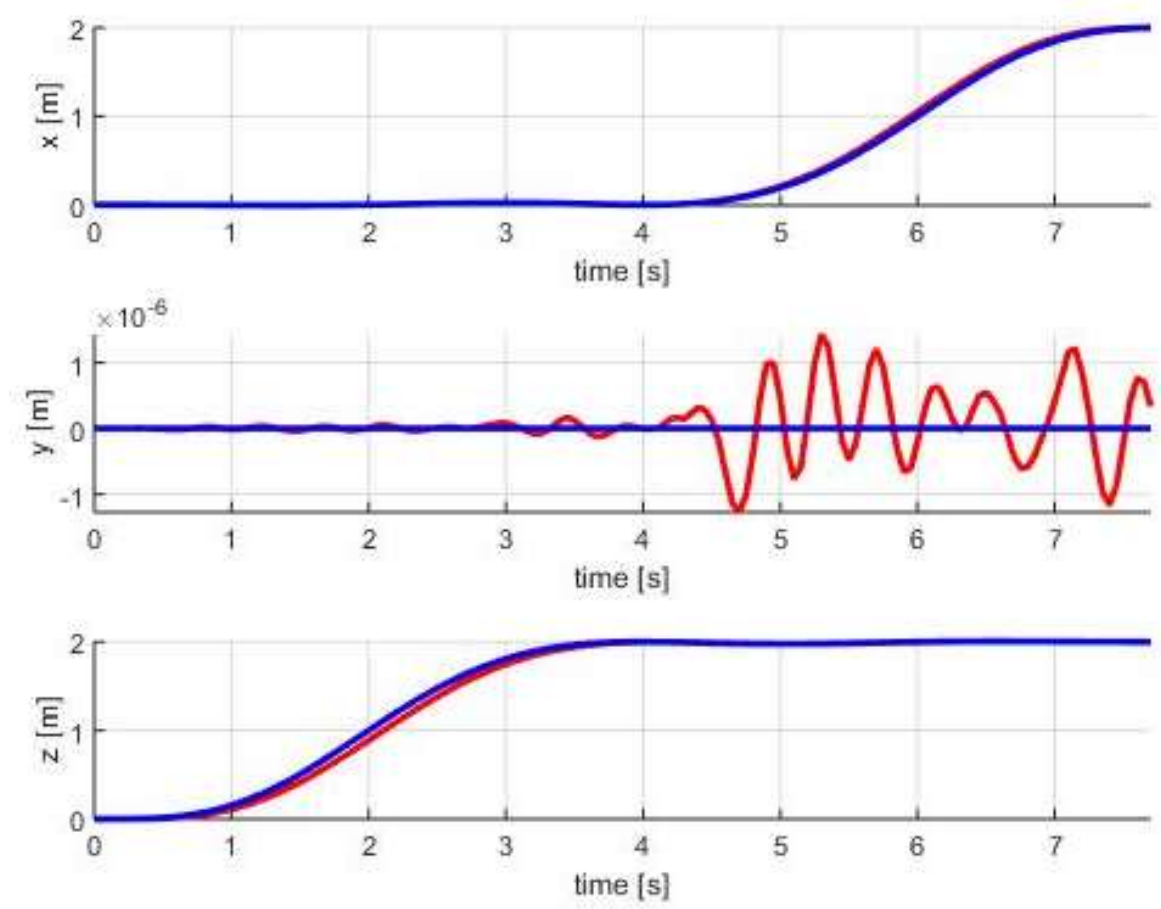

Figure 3.1. Path generated

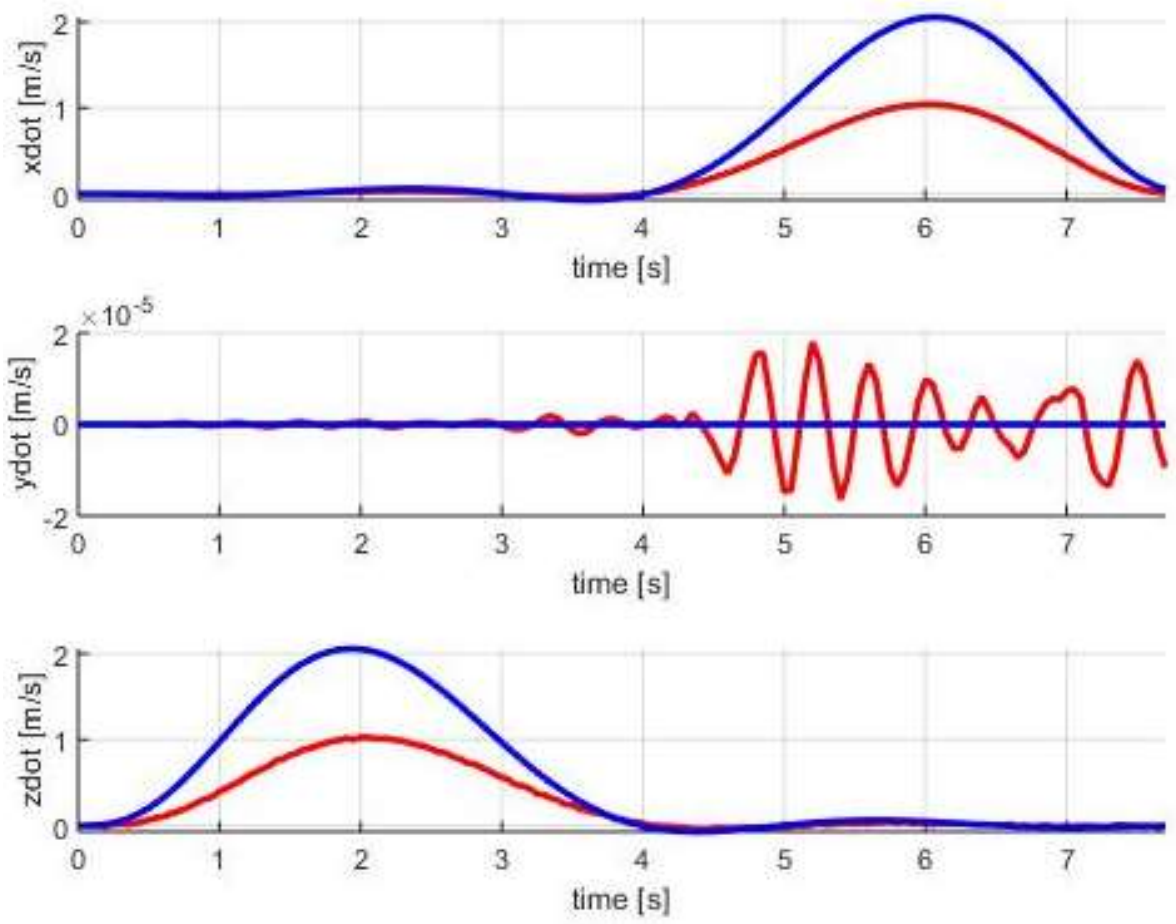

Figure 3.2. Velocity generated

# Eigenvalue Picking Method

By assigning a moving reference, the difference would decrease. Consequently, the control signal decrease. However, it increases the difficulty in designing the controller; the eigenvalues can be easily picked with specific requirement when the reference is fixed at the target position only.

Developing the standard method in picking a proper eigenvalue with specific requirement is demanded. The method in picking eigenvalue with a moving reference has been created for the first time in this study.

We will introduce this method with a real problem:

**Assumption 1.** The reference is moving from 0 to L with uniform acceleration where the initial velocity is 0, the velocity at final position is 2m/s. The total moving time is 2s. The velocity-time relationship has been plotted in Figure 3.3.

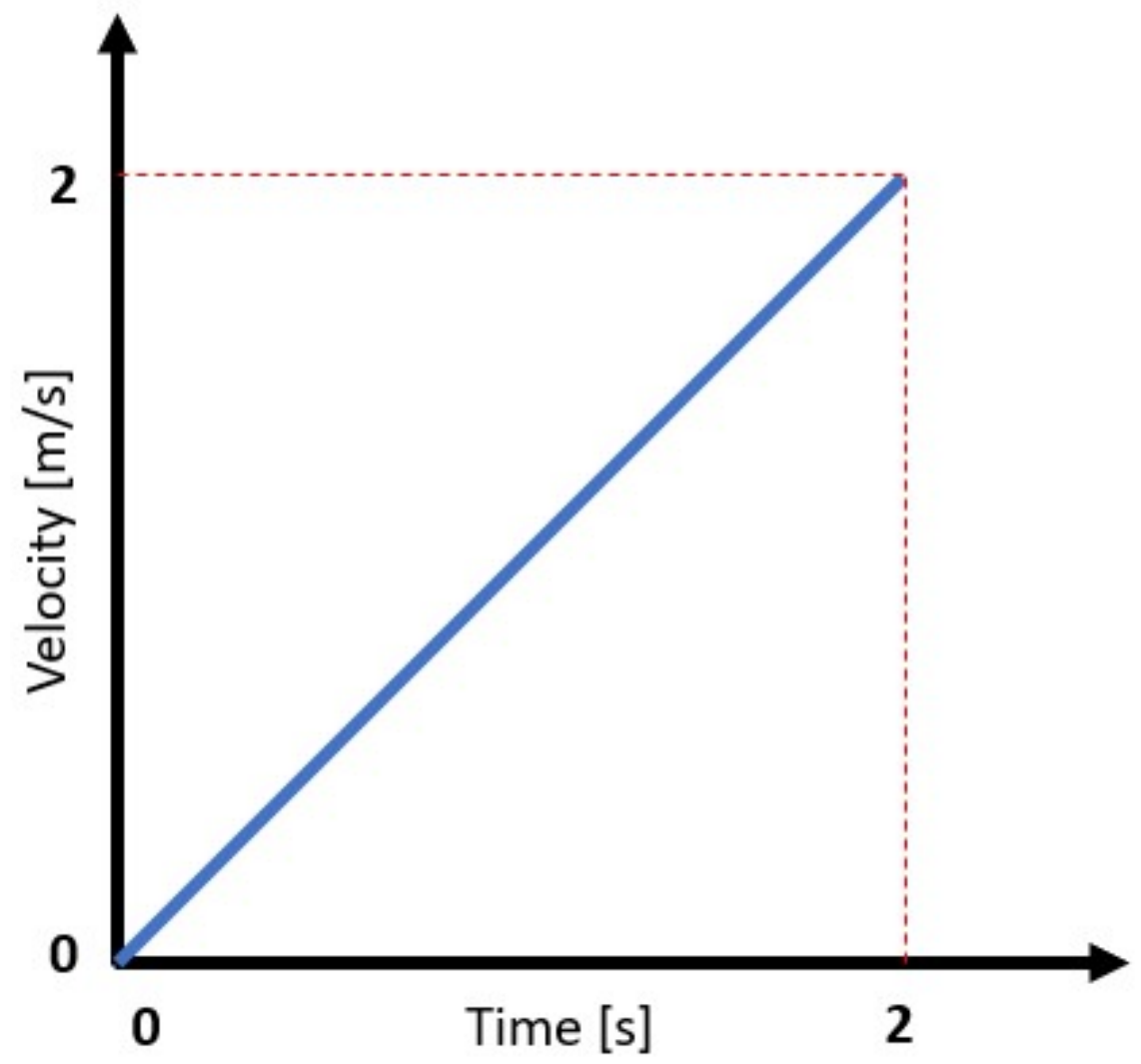

Figure 3.3. The v-t plot of the moving reference

**Assumption 2.** The controller used here is a first order controller. This controller makes the accomplishment of movement $p = 1 - e^{\lambda \cdot t_{time}}$ at $t_{time}$ if the reference is fixed. Here $\lambda$ is the eigenvalue (negative). When p=100%, the object is fully controlled.

**Question.** Where is the object when the reference reaches the target L (t = 2s)?

**Answer:**

Step 1. Equally divide the total moving time (2s) into n pieces. So that each segment lasts $\frac{t}{n}$ second. Assume that the distance the reference moves during each time segment is $x_1$, $x_2$, $x_3$,…, $x_n$. (Fig.3.4)

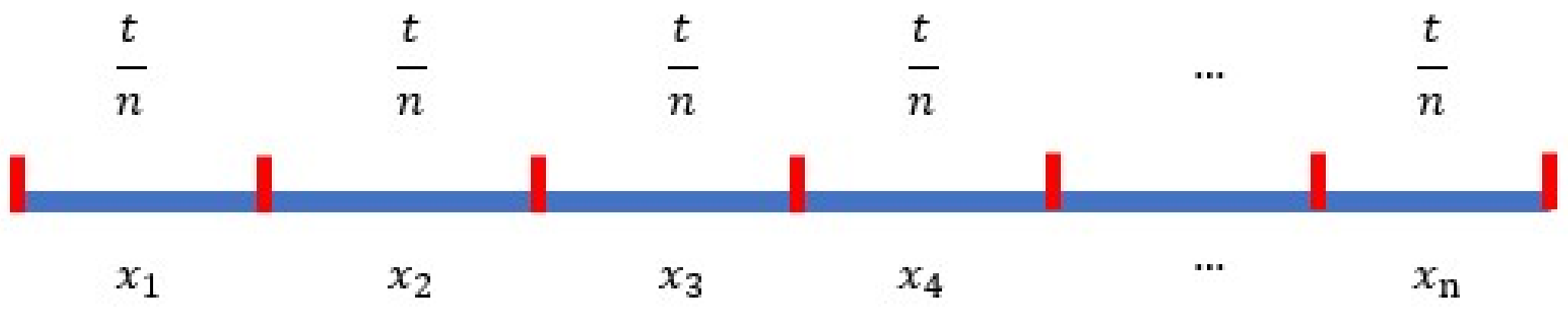

Figure 3.4. L is divided into n pieces. In each piece the time is $\frac{t}{n}$.

Step 2. Assume the reference is fixed at the end position of each segment (red dots in Fig.3.4) during each time segment.

It means that the reference is fixed at point $A_1$ during the first time-segment lasting $\frac{t}{n}$. And the reference switched to $A_2$ at time $\frac{t}{n}$. Then it keeps being fixed at $A_2$ for $\frac{t}{n}$ before switching to $A_3$…until the reference reaches $A_n$. (Fig.3.5)

We may also call it a time-equal step-reference serial.

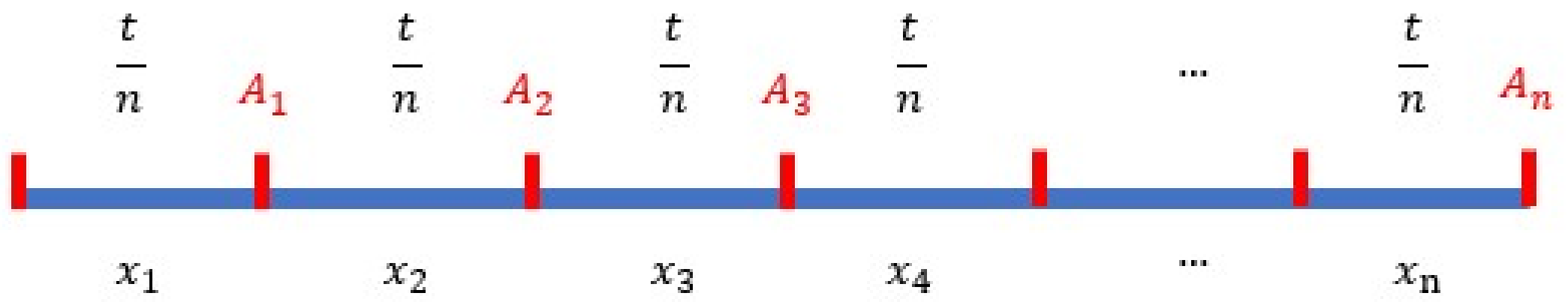

Figure 3.5. The reference is switched forward to each point located at the position end of each time-segment

Step 3. Calculate the actual moving distance of the object during each time-segment. Notice $p = 1 - e^{\lambda \cdot \frac{t}{n}}$ is same in each time-segment.

During the first time-segment, the actual moving distance is

$$x_1 \cdot p$$

During the second time-segment, the actual moving distance is

$$(x_1 + x_2 - x_1 \cdot p) \cdot p$$

During the third time-segment, the actual moving distance is

$$(x_1 + x_2 + x_3 - 2\, x_1 \cdot p - x_2 \cdot p + x_1 \cdot p^2) \cdot p$$

…

Here are all the terms in all of the time-segments:

$$\begin{array}{ccccc} x_1 & x_2 & x_3 & \cdots & x_n \\ x_1 \cdot p & x_2 \cdot p & x_3 \cdot p & \cdots & x_{n-1} \cdot p \\ x_1 \cdot p^2 & x_2 \cdot p^2 & x_3 \cdot p^2 & \cdots & x_{n-2} \cdot p^2 \\ & & \cdots & & \\ x_1 \cdot p^{n-1} & & & & \end{array}$$

The sum of the coefficient of $x_i$ can be calculated based on Pascal Triangle (Fig.3.6).

1

1 1

-1 2 1

1 -3 3 1

-1 4 -6 4 1

1 -5 10 -10 5 1

Figure 3.6 Pascal's Triangle

Step 4. Calculate the total moving distance of the object by adding the moving distance during each segment together.

The final sum of coefficient of $\frac{x_1}{p}$ accumulated from Pascal's Triangle is collected below (3.1). We need to try to simplify (3.1). And the process of it is discussed below in (3.1) to (3.5).

$$
\begin{aligned}
& n \\
& - p * (1 + 2 + 3 + \cdots + n - 1) \\
& + p^2 * (C_2^2 + C_3^2 + C_4^2 + C_5^2 + \cdots + C_{n-1}^2) \\
& - p^3 * (C_3^3 + C_4^3 + C_5^3 + C_6^3 + \cdots + C_{n-1}^3) \\
& + \cdots \\
& - p^{n-1} * C_{n-1}^{n-1}
\end{aligned}
\tag{3.1}
$$

Term (3.1) can be rewritten in (3.2) if we extract a negative symbol.

$$
\begin{aligned}
&1 + (n-1) \\
&\quad + (-p)^1 * (1 + 2 + 3 + \cdots + n - 1) \\
&\quad + (-p)^2 * (C_2^2 + C_3^2 + C_4^2 + C_5^2 + \cdots + C_{n-1}^2) \\
&\quad + (-p)^3 * (C_3^3 + C_4^3 + C_5^3 + C_6^3 + \cdots + C_{n-1}^3) \\
&\quad + \cdots \\
&\quad + (-p)^{n-1} * C_{n-1}^{n-1}
\end{aligned}
$$

(3.2)

(3.2) can be rewritten in (3.3) if we manually create $(-p)^0$

$$
\begin{aligned}
&1 + (-p)^0 * (C_1^0 + C_2^0 + C_3^0 + C_4^0 + \cdots + C_{n-1}^0) \\
&\quad + (-p)^1 * (C_1^1 + C_2^1 + C_3^1 + C_4^1 + \cdots + C_{n-1}^1) \\
&\quad + (-p)^2 * (C_2^2 + C_3^2 + C_4^2 + C_5^2 + \cdots + C_{n-1}^2) \\
&\quad + (-p)^3 * (C_3^3 + C_4^3 + C_5^3 + C_6^3 + \cdots + C_{n-1}^3) \\
&\quad + \cdots \\
&\quad + (-p)^{n-1} * C_{n-1}^{n-1}
\end{aligned}
$$

(3.3)

The result of (3.3) is recalculated based on Binomial Theorem. And the result yields:

$$1 + (1-p)^1 + (1-p)^2 + \cdots + (1-p)^{n-1} \quad (3.4)$$

(3.4) is the sum of a geometric progression. The sum of it is listed in (3.5).

$$\frac{1-(1-p)^n}{p} \quad (3.5)$$

So that the coefficient of $x_1$ is $1-(1-p)^n$ . Similarly, the coefficients of $x_2, x_3, \dots, x_n$ can be calculated. So far, the detail of the method in calculating the sum of each $x_i$ has been illustrated in (3.1) to (3.5).

Notice $x_1 + x_2 + x_3 + \cdots + x_n = L$, the result of total moving distance of the object is

$$L - [x_1 \cdot (1-p)^n + x_2 \cdot (1-p)^{n-1} + x_3 \cdot (1-p)^{n-2} + \cdots + x_n \cdot (1-p)^1]$$

If n is large, the time-segment is little enough so that the velocity during each time-segment can be regarded constant. In other words, the previous result can be rewritten as

$$L - \frac{t}{n} \cdot [v_1 \cdot (1-p)^n + v_2 \cdot (1-p)^{n-1} + v_3 \cdot (1-p)^{n-2} + \cdots + v_n \cdot (1-p)^1]$$

(3.5)

Step 5. Seek the limit of (3.5) when $n = \infty$

As what we have done in previous steps, the continuous time are divided into the combination of infinite time-pieces. Consequently, the result of the limit in Step 5 is exactly the final answer to the question.

We are to find the limit of the following part first:

$$\frac{1}{n} \cdot [v_1 \cdot (1-p)^n + v_2 \cdot (1-p)^{n-1} + v_3 \cdot (1-p)^{n-2} + \cdots + v_n \cdot (1-p)^1]$$

(3.6)

If we pick the eigenvalue $\lambda = -10$, we have

$$p = 1 - e^{\frac{-20}{n}} \quad (3.7)$$

Also notice that we have

$$v_i = \frac{2}{n} \cdot i \quad (3.8)$$

Substitute (3.7) and (3.8) into (3.6), the result yields:

$$\frac{2}{n^2} \cdot \left[1 \cdot e^{-\frac{20}{n} \cdot n} + 2 \cdot e^{-\frac{20}{n} \cdot (n-1)} + 3 \cdot e^{-\frac{20}{n} \cdot (n-2)} + \cdots + n \cdot e^{-\frac{20}{n} \cdot 1}\right]$$

(3.9)

To simplify (3.9), first we define the sum in bracket of (3.9) $S_n$:

$$S_n = 1 \cdot e^{-\frac{20}{n} \cdot n} + 2 \cdot e^{-\frac{20}{n} \cdot (n-1)} + 3 \cdot e^{-\frac{20}{n} \cdot (n-2)} + \cdots + n \cdot e^{-\frac{20}{n} \cdot 1}$$

(3.10)

Multiply $e^{\frac{20}{n}}$ both sides in (3.10) yields:

$$e^{\frac{20}{n}} \cdot S_n = 1 \cdot e^{-\frac{20}{n} \cdot (n-1)} + 2 \cdot e^{-\frac{20}{n} \cdot (n-2)} + 3 \cdot e^{-\frac{20}{n} \cdot (n-3)} + \cdots + n \cdot e^{-\frac{20}{n} \cdot 0}$$

(3.11)

(3.10) minus (3.11) yields:

$$\left(1 - e^{\frac{20}{n}}\right) \cdot S_n = e^{-\frac{20}{n} \cdot n} + e^{-\frac{20}{n} \cdot (n-1)} + e^{-\frac{20}{n} \cdot (n-2)} + e^{-\frac{20}{n} \cdot (n-3)} + \cdots + e^{-\frac{20}{n} \cdot 0}$$

(3.12)

The right of equation (3.12) can now be calculated; it is the sum of equal ratio series. Consequently, $S_n$ can be calculated. The result is:

$$S_n = \frac{e^{-20} - 1}{\left(1 - e^{\frac{20}{n}}\right)^2} - \frac{n}{1 - e^{\frac{20}{n}}} \quad (3.13)$$

Substitute (3.9), (3.10) and (3.13) into (3.5). And get the limit using Lobita's Law. The result yields

L – 0.1900 (3.14)

So, L – 0.1900 is the final answer if we pick the eigenvalue -10.

It means that the total moving distance of the object is L–0.1900 with the assumptions that the reference is moving follow v-t plot in Fig.3.3 and the controller is a first order controller with the eigenvalue -10.

This result (3.14) will be used in further discussion in altitude control in Chapter 'Basic Control'.

Furthermore, this eigenvalue picking method or process can be used in any cases where reference is moving. (3.5) is true without special requirement of the controller or moving pattern of reference. In other words, the controller is not required to be first order. And the reference can move in an irregular way. (3.5) always provides the accomplishment of movement accurately.

# CHAPTER 4

# Basic Control

The statement 'to get a quadrotor well-controlled' is equivalent to the statement 'get the altitude and the attitude controlled'. Here, altitude represents the height of the quadrotor. The attitude represents the roll, pitch and yaw of the quadrotor. The controller is to be designed in this chapter.

## Altitude Controller

Consider the equation (2.2):

$$\begin{pmatrix} \dot{z} \\ \ddot{z} \end{pmatrix} = \begin{bmatrix} 0 & 1 \\ 0 & 0 \end{bmatrix} \begin{pmatrix} z \\ \dot{z} \end{pmatrix} + \begin{bmatrix} 0 & 0 & 0 & 0 \\ \frac{1}{m} & \frac{1}{m} & \frac{1}{m} & \frac{1}{m} \end{bmatrix} \begin{pmatrix} F_1 \\ F_2 \\ F_3 \\ F_4 \end{pmatrix} + \begin{pmatrix} 0 \\ -g \end{pmatrix}$$

or $\dot{q} = A \cdot q + B \cdot u + v$

The feedback is introduced here by picking a proper input u.

We write the u in the following form

$$u = -k \cdot q + N \cdot r \tag{4.1}$$

Here, k determines the controlling speed. N determines the state after being stable. r is a 2 by 1 reference vector.

Before moving on, we will make the requirements for the controller:

Requirement 1: State q shall chase reference vector r.

Requirement 2: State q should reach at least 80% of the target after 2 seconds.

To meet the requirement 1, we are to pick a proper N. The process of it is discussed below.

Substitute equation (4.1) into the state space equation (2.2), we have:

$$\dot{q} - \mathrm{A} \cdot \mathrm{q} + \mathrm{B} \cdot \mathrm{k} \cdot \mathrm{q} - \mathrm{B} \cdot \mathrm{N} \cdot \mathrm{r} = \mathrm{v} \tag{4.2}$$

Consider the following relationship:

$$[\,1 \quad 0\,] \cdot \dot{q} \;=\; [\,0 \quad 1\,] \cdot q \tag{4.3}$$

It is true because both left side and right side equal to $\dot{z}$.

Moreover,

$$\dot{q} = \; pinv([\,1 \quad 0\,]) \cdot [\,0 \quad 1\,] \cdot q \tag{4.4}$$

Notice r is the final state:

$$r = \begin{pmatrix} h \\ 0 \end{pmatrix}$$

h is the desired height of the quadrotor.

So that after being stable

$$\mathrm{q} = \mathrm{r} \tag{4.5}$$

Substitute (4.4) and (4.5) into (4.2). The result yields:

$$\mathrm{B} \cdot (\,\mathrm{k} \; - \mathrm{N}\,) \cdot \begin{pmatrix} \mathrm{h} \\ 0 \end{pmatrix} = \mathrm{v} \tag{4.6}$$

We can rewrite the result in a more detailed form:

$$N_{11} + N_{21} + N_{31} + N_{41} = k_{11} + k_{21} + k_{31} + k_{41} + \frac{1}{\mathrm{h}} * mg \tag{4.7}$$

If we write N and K in detail:

$$N = \begin{pmatrix} \mathrm{N}_{11} & \mathrm{N}_{12} \\ \mathrm{N}_{21} & \mathrm{N}_{22} \\ \mathrm{N}_{31} & \mathrm{N}_{32} \\ \mathrm{N}_{41} & \mathrm{N}_{42} \end{pmatrix} \qquad \mathrm{K} = \begin{pmatrix} \mathrm{k}_{11} & \mathrm{k}_{12} \\ \mathrm{k}_{21} & \mathrm{k}_{22} \\ \mathrm{k}_{31} & \mathrm{k}_{32} \\ \mathrm{k}_{41} & \mathrm{k}_{42} \end{pmatrix}$$

We call this relationship **Chasing Condition**. Once the N and K meet the Chasing Condition, state after being stable will be exact to reference.

To meet the requirement 2, we are to pick a proper k by selecting a set of proper eigenvalues (note there will be 2 eigenvalues since it is a second order system).

Before that, we need to reanalyze the height reference generated in path planning (Fig.4.1) and the velocity reference generated (Fig.4.2).

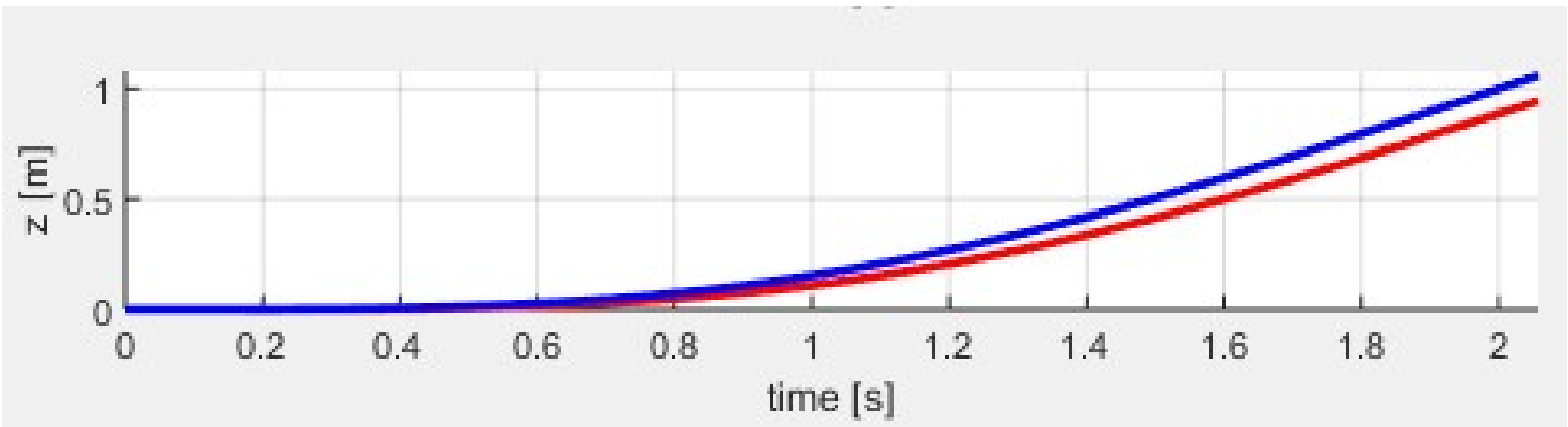

Figure 4.1. The blue line is the altitude reference generated in path planning

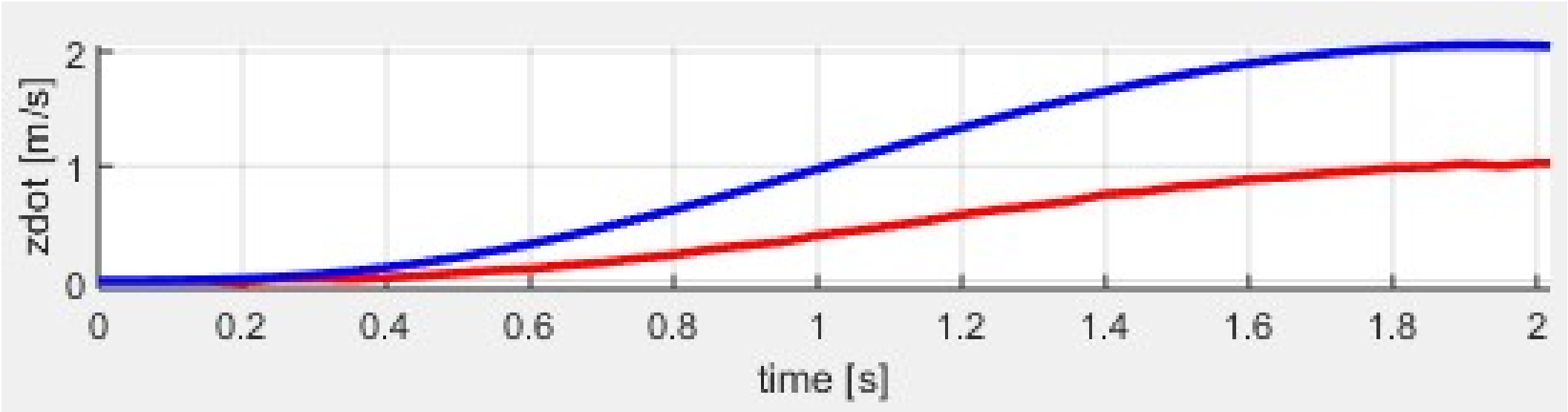

Figure 4.2. The blue line is the velocity reference generated in path planning

The designed reference moves from 0 to 1 meter high with an increasing velocity from 0 to 2 m/s. In other words, it is a controlling problem where the reference is not fixed. Whereas, this problem is thoroughly discussed in 'Eigenvalue Picking Method' in previous chapter 'Path Planning'.

There are 2 differences between the discussed example in 'Eigenvalue Picking Method' and the case here. One is that the reference is not moving in a constant velocity. It can be judged from Fig.4.2 that the velocity is now moving in a smoother way. The other is that the quadrotor is a second order system. While the result deduced in 'Eigenvalue Picking Method' based on the assumption that the system is a first-order one.

Despite these differences, there are still a lot of similar characters.

Firstly, both velocities are increasing in a similar tendency. It can be deduced from Fig. 3.3 and Fig.4.2. Additionally, both times we interested in are 0 – 2s where the velocity accelerates. So, based on these two similarities, we assume that the real velocity reference generated can be roughly regarded as the case discussed in 'Eigenvalue Picking Method'.

Since we have not discussed the way to pick eigenvalues for second-order system. We pick one dominant eigenvalue which is 0.1 times to the other eigenvalue in value (e.g. eigenvalue pairs -10 and -100). The behavior of the system with this eigenvalue pair is similar to the first order system.

By picking the eigenvalue pair this way, the eigenvalue picking problem in altitude controller designing in this chapter can use the method we developed in 'Eigenvalue Picking Method'. And the example we discussed in 'Eigenvalue Picking Method' is the very question we are to solve; the result (3.14) can be used here.

When t = 2s, the reference planned moves to 1 meter (Fig.4.1). We substitute L=1 into (3.14). It yields 0.8100.

**It means that the quadrotor is supposed to move to 81.00% of the reference after 2 second if we pick -10 as the eigenvalue with the assumption that the quadrotor is first order system and the planned velocity increases constantly.**

As we discussed, 81.00% is a rough estimation of the controlling result if we pick eigenvalue pair ( -100 , -10 ). It meets the requirement 2. And [ -100  -10 ] is exactly the eigenvalue pair we used in controlling the height.

Using MATLAB function 'place', the k in (4.1) is determined. Here is the function we used in MATLAB:

K = place( A, B, [ -100  -10 ] )

As we mentioned, the N is determined based on equation (4.7) to make sure that the height will reach the reference after being stable.

Note that (4.7) only defines the sum of the first column of N. In my research, each of the element in the first column is set as the same value to each other; it can be done by dividing 4 by the sum. The undefined 4 elements in the second column are set 1.

Since k and N have been well-defined, the u is now achieved using equation (4.1). This is the input we need for controlling altitude/height.

## Attitude Controller

The roll pitch and yaw are to be controlled in this section.

In general, a PD controller is applied to control the attitude.

$$M = k_p \cdot \left[ \begin{pmatrix} \varphi_{desired} \\ \theta_{desired} \\ \psi_{desired} \end{pmatrix} - \begin{pmatrix} \varphi \\ \theta \\ \psi \end{pmatrix} \right] + k_d \cdot \left[ \begin{pmatrix} p_{desired} \\ q_{desired} \\ r_{desired} \end{pmatrix} - \begin{pmatrix} p \\ q \\ r \end{pmatrix} \right] \tag{4.8}$$

Here, $\varphi$, $\theta$, and $\psi$ are roll pitch and yaw of the quadrotor. p, q, and r are angular velocity along each axis of the fixed frame. The result M is the moment needed to control the attitude.

Here, we need to determine 6 desired states, $\varphi_{desired}$, $\theta_{desired}$, $\psi_{desired}$, $p_{desired}$, $q_{desired}$, and $r_{desired}$, first.

Since we don't want to set a rotating state as desired state:

$$p_{desired} = 0 \tag{4.9}$$

$$q_{desired} = 0 \tag{4.10}$$

$$r_{desired} = 0 \tag{4.11}$$

Also, we expect 0 rad in yaw:

$$\psi_{desired} = 0 \tag{4.12}$$

We are going to determine $\varphi_{desired}$ and $\theta_{desired}$ next.

Consider the equation (2.3)

$$m\,\ddot{r} = \begin{pmatrix} 0 \\ 0 \\ -mg \end{pmatrix} + R\begin{pmatrix} 0 \\ 0 \\ F_1 + F_2 + F_3 + F_4 \end{pmatrix}$$

The raw ($\varphi$) and pitch ($\theta$) are little (usually less than 0.1) in our movement, so that we have the following estimations

$$\sin\theta = \theta \quad \cos\theta = 1 \tag{4.13}$$

$$\sin\varphi = \varphi \quad \cos\varphi = 1 \tag{4.14}$$

Additionally, we assume:

$$F_1 + F_2 + F_3 + F_4 = mg \tag{4.15}$$

$$\psi = \psi_0 \tag{4.16}$$

Substitute (4.9) to (4.16) into equation (2.3). The result of the first 2 lines yields:

$$\ddot{r_1} = g \cdot (\theta \cdot \cos\psi_0 + \varphi \cdot \sin\psi_0) \tag{4.17}$$

$$\ddot{r_2} = g \cdot (\theta \cdot \sin\psi_0 - \varphi \cdot \cos\psi_0) \tag{4.18}$$

We can get the expressions of $\theta$ and $\varphi$ from (4.17) and (4.18):

$$\varphi = \frac{1}{g} \cdot (\ddot{r_1} \cdot \sin\psi_0 - \ddot{r_2} \cdot \cos\psi_0) \tag{4.19}$$

$$\theta = \frac{1}{g} \cdot (\ddot{r_1} \cdot \cos\psi_0 + \ddot{r_2} \cdot \sin\psi_0) \tag{4.20}$$

$\ddot{r_1}$ and $\ddot{r_2}$ are the accelerations along x and y axis (world-fixed frame). It gives a relationship between the accelerations and $\varphi$, $\theta$.

The desired accelerations have been generated from path planning. So that the desired attitude, $\varphi_{desired}$ and $\theta_{desired}$, can be calculated from (4.19) and (4.20).

So far, all of the 6 desired states, $\varphi_{desired}$, $\theta_{desired}$, $\psi_{desired}$, $p_{desired}$, $q_{desired}$, and $r_{desired}$, have been explained.

The coefficients $k_p$ and $k_d$ are picked by trying to make the quadrotor not to flip mistakenly in moving. And in my thesis, the $k_p$ and $k_d$ are selected as:

$$k_p = [\,200\,;200\,;500\,]$$

$$k_d = [\,10\,;10\,;10\,]$$

Now, the moment for controlling attitude can be calculated from (4.8). However, the thrust is the only input. So, the moment is translated to the thrust generating this moment. The relationship between moment M and thrusts is:

$$M = \begin{bmatrix} 0 & L & 0 & -L \\ -L & 0 & L & 0 \\ r & -r & r & -r \end{bmatrix} \begin{pmatrix} F_1 \\ F_2 \\ F_3 \\ F_4 \end{pmatrix} \tag{4.21}$$

So that the thrusts can be achieved:

$$\begin{pmatrix} F_1 \\ F_2 \\ F_3 \\ F_4 \end{pmatrix} = pinv\left( \begin{bmatrix} 0 & L & 0 & -L \\ -L & 0 & L & 0 \\ r & -r & r & -r \end{bmatrix} \right) \cdot M \tag{4.22}$$

Here, 'pinv' represents the pseudo inverse.

The thrusts set we achieved here is the input we need to control the attitude.

## Final Input

In controlling the altitude/height, we got a desired input $u_{altitude}$. Also, in controlling the attitude, we got another desired input $u_{attitude}$.

We need to modify or combine these two control signals to control the altitude and attitude simultaneously.

There are 2 methods for this.

One is by inputting different kinds of input at different time by a switch. (Figure 4.3)

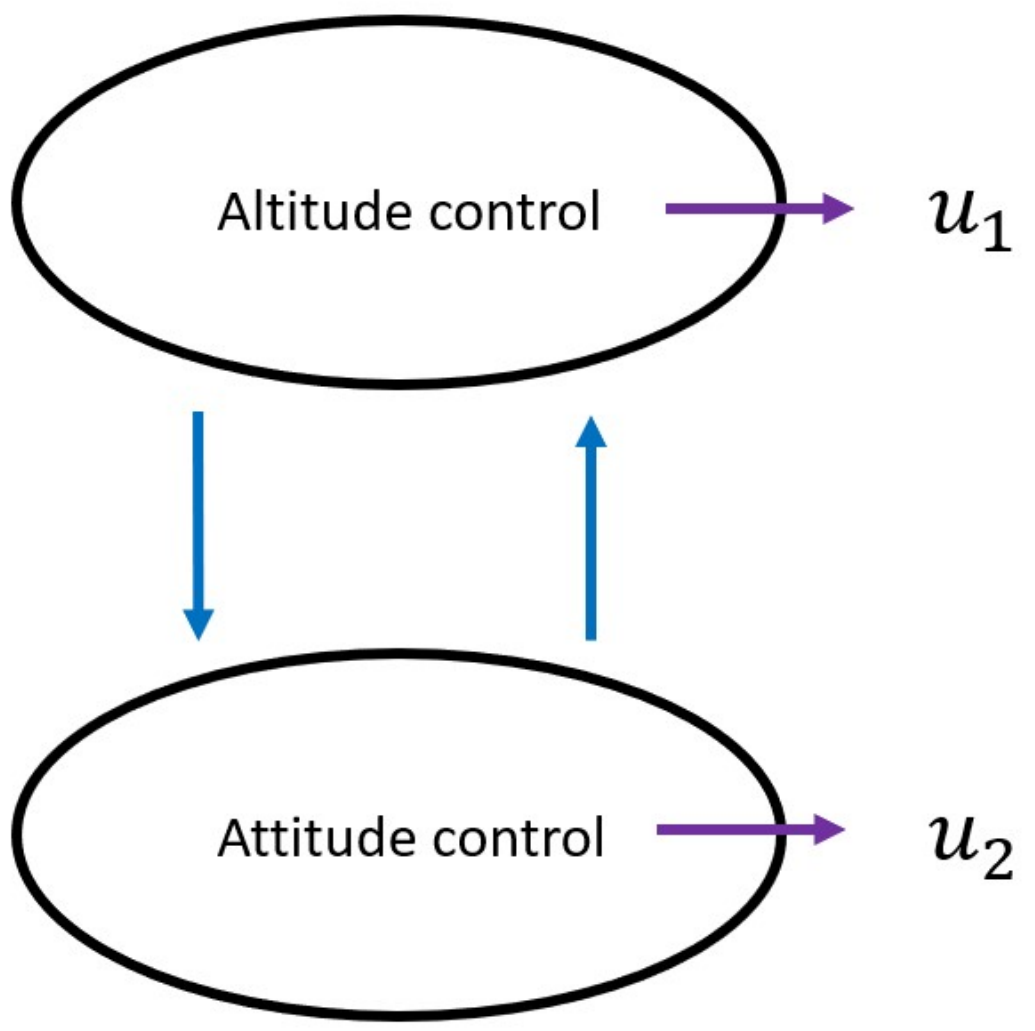

Figure 4.3. The category of input changes in need for different cases

A detecting system with a threshold tells the controller which state the quadrotor is under. If the quadrotor is under the situation with large tendency to flip (large angular acceleration) unproperly, the input is $u_{attitude}$ only. If the quadrotor is under the situation with large tendency to lift or fall (large acceleration) unproperly, the input is $u_{altitude}$ only. These two states are switching frequently in a flight.

However, the motor controlling demands time. Switching too frequently may cause the large delay in controlling. So, we may need another method.

The actual method of combining signal we used in quadrotor control in my research is by simply adding them.

$$u_{final} = u_{altitude} + u_{attitude}$$

The sum of control signals in controlling attitude and control signal in controlling altitude is the final input/control signal in quadrotor controlling.

The result is to be verified in the next chapter 'Simulator and Result'.

# CHAPTER 5

# Simulator and Result

In this chapter, the environment and mechanism of the simulator used to verify the designed controller will be introduced. After that, the simulating results with the designed controller will be plotted.

## Simulator

The Simulator is written in MATLAB m files. It mainly consists of 4 parts, namely solver, trajectory generator/path planning, controller, and output defining.

The input is the points that the quadrotor is supposed to pass. In our case, the input is defined as the following points matrix (5.1).

$$\begin{bmatrix} 0 & 0 & 0 \\ 0 & 0 & 1 \\ 0 & 0 & 2 \\ 1 & 0 & 2 \\ 2 & 0 & 2 \end{bmatrix} \tag{5.1}$$

The quadrotor is to lift from [0 0 0]. After pass [0 0 1], it goes to [0 0 2]. And then it goes another direction to pass [1 0 2] before a final reach to [2 0 2].

The target is to control the quadrotor lifting from [0 0 0] to [0 0 2] before translating to [2 0 2]. In (5.1), however, two extra waypoints, [0 0 1] and [1 0 2], are also listed. The reason for this relates to the algorithm of path planning.

In path planning, our generated path is as smooth as possible in our algorithm mentioned. If we define the waypoints as follow (5.2).

$$\begin{bmatrix} 0 & 0 & 0 \\ 0 & 0 & 2 \\ 2 & 0 & 2 \end{bmatrix} \tag{5.2}$$

The reference will pass [0 0 2], while the path near the turning points will bias a lot from the line linking [0 0 0] to [0 0 2]. This is the adverse effect of our smooth-reference-designing algorithm. If we add a passing point [0 0 1], it will deter this bias to some extent.

The output consists of 14 figures recording the states during flying. They are real flying animation, position, velocity, thrusts, roll, pitch, yaw, $r_1$, $r_2$, $r_3$, $r_4$, p, q, and r. $r_1$, $r_2$, $r_3$, and $r_4$ are the ratios of M to F we discussed in (2.5). p, q, and r are the angular velocities in the direction of each axis with respective to the body-fixed frame.

The mechanism of the simulator will be discussed here. The core part is a function to solve two ordinary differential equations. The equations are:

$$m\,\ddot{r} = \begin{pmatrix} 0 \\ 0 \\ -mg \end{pmatrix} + R\begin{pmatrix} 0 \\ 0 \\ F_1 + F_2 + F_3 + F_4 \end{pmatrix}$$

$$I\begin{pmatrix} \dot{p} \\ \dot{q} \\ \dot{r} \end{pmatrix} = \begin{bmatrix} 0 & L & 0 & -L \\ -L & 0 & L & 0 \\ r & -r & r & -r \end{bmatrix}\begin{pmatrix} F_1 \\ F_2 \\ F_3 \\ F_4 \end{pmatrix} - \begin{bmatrix} p \\ q \\ r \end{bmatrix} \times I\begin{bmatrix} p \\ q \\ r \end{bmatrix}$$

These equations have already been introduced in (2.3) and (2.5). It can also be seen that we established the right dynamic equation to design the controller.

The solver is ODE45. It selects the step time automatically. However, we only output the result at some specific time points; the results at each 0.05 second are output and used in the plot only. But it doesn’t mean that the calculation only happens in these time points. The actual time points for calculating is determined by ODE45 itself.

# Result and Analyze

In this section, the simulating result with our controller will be displayed before being analyzed. The controlling parameters have been introduced in Chap 4.

As mentioned, the input is set as matrix (5.1). The followings are the outputs.

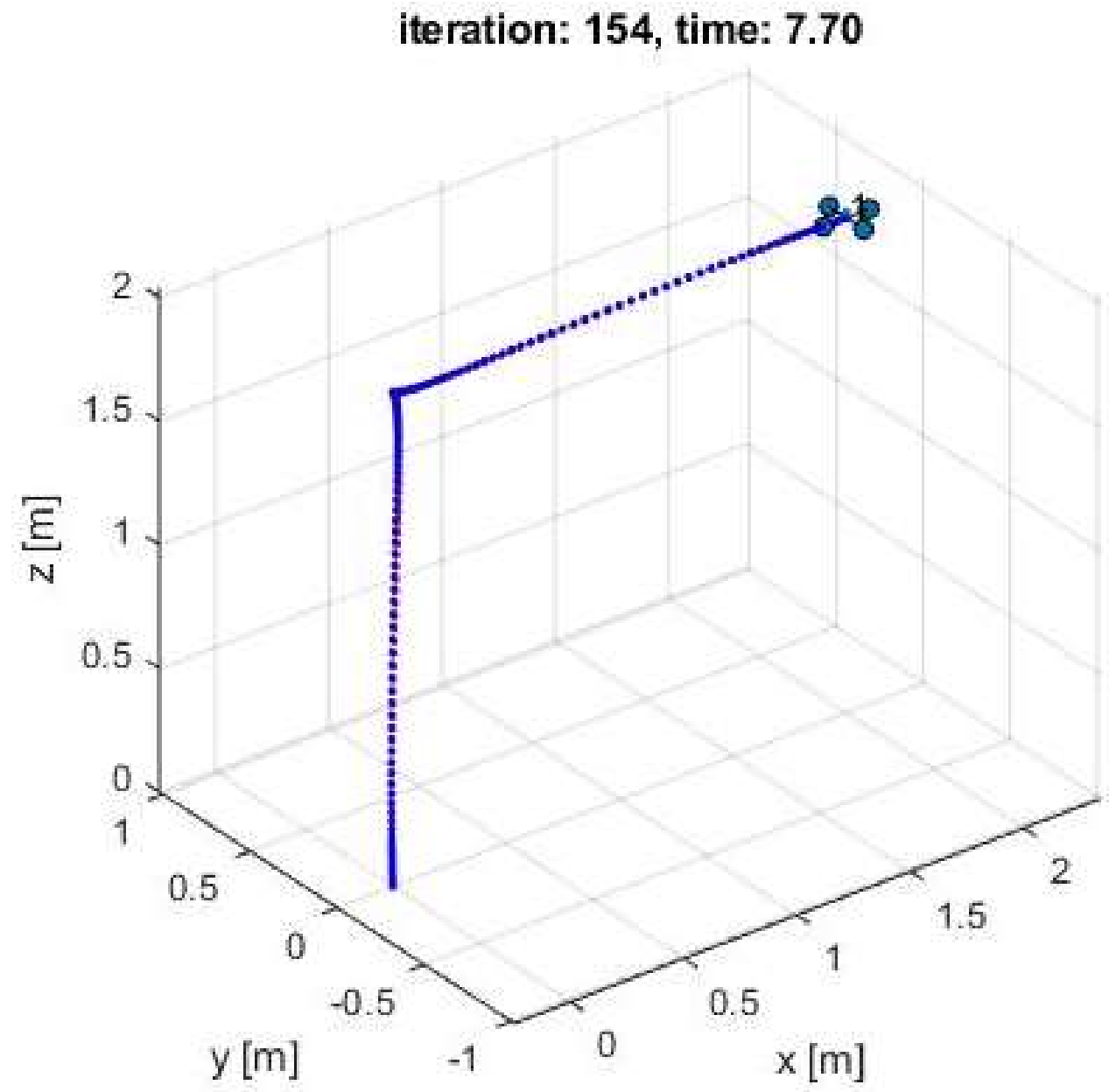

Figure 5.1. The animation of the quadrotor

It can be seen from Fig.5.1 that there are total 7.7 seconds for quadrotor to move. And it is within 154 iterations/outputs; simulator outputs data every 0.05 second.

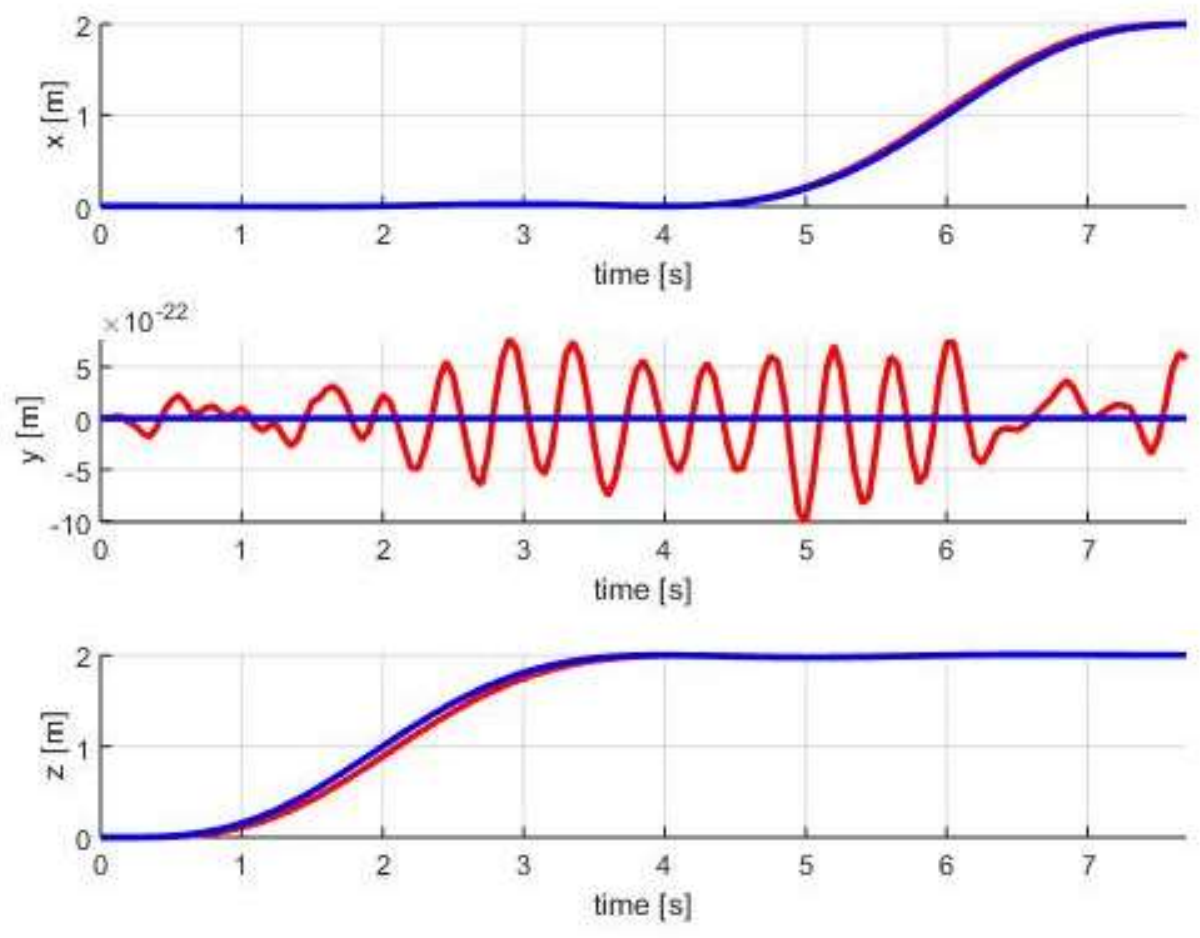

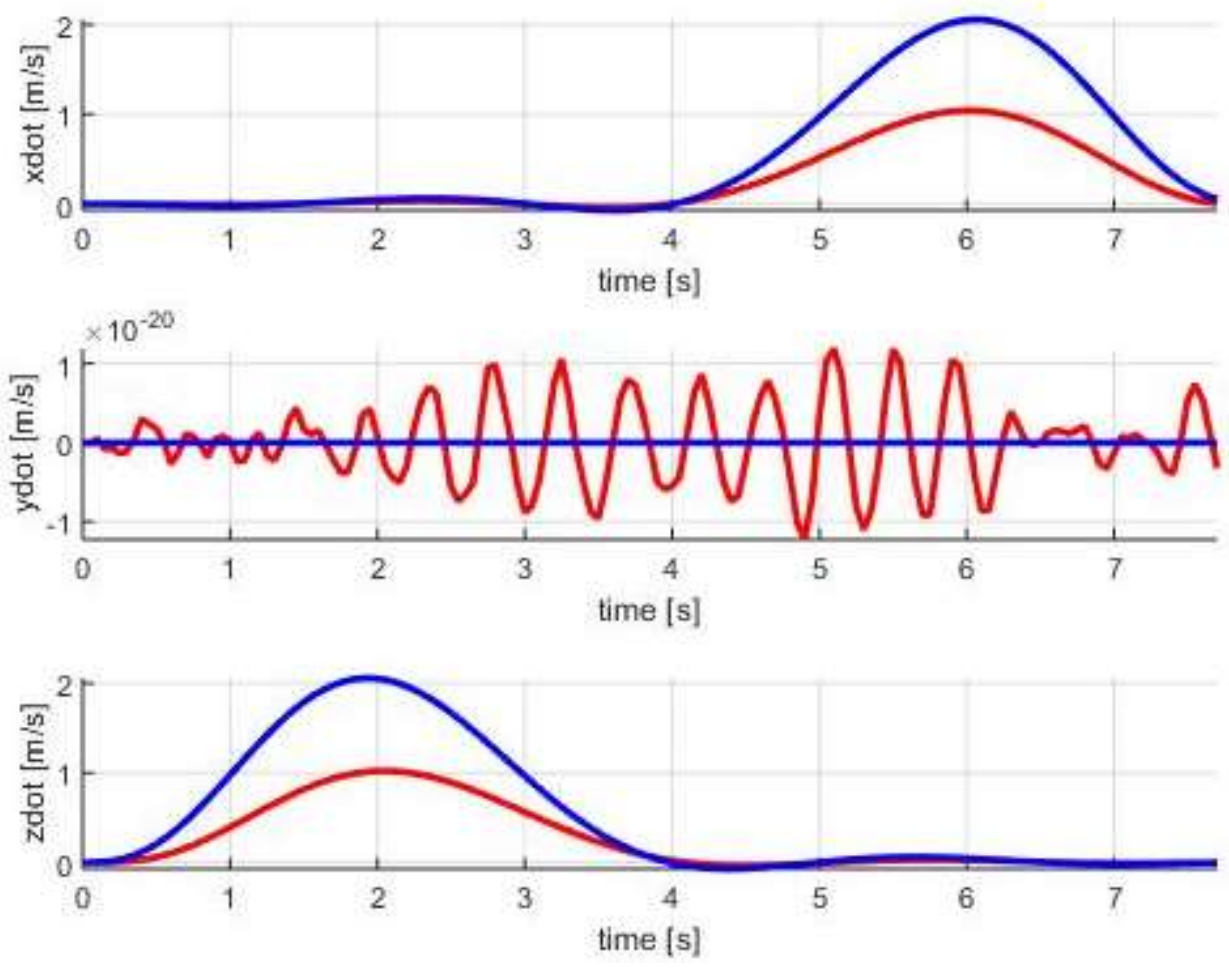

Figure 5.2. Position of the quadrotor

Figure 5.3. Velocity of the quadrotor

The position and the velocity of the quadrotor outputs are plotted in Fig.5.2. and Fig.5.3. These values are displayed in 3 dimensions, x, y, and z. The blue line is the reference generated in path planning. The red line is the actual moving information.

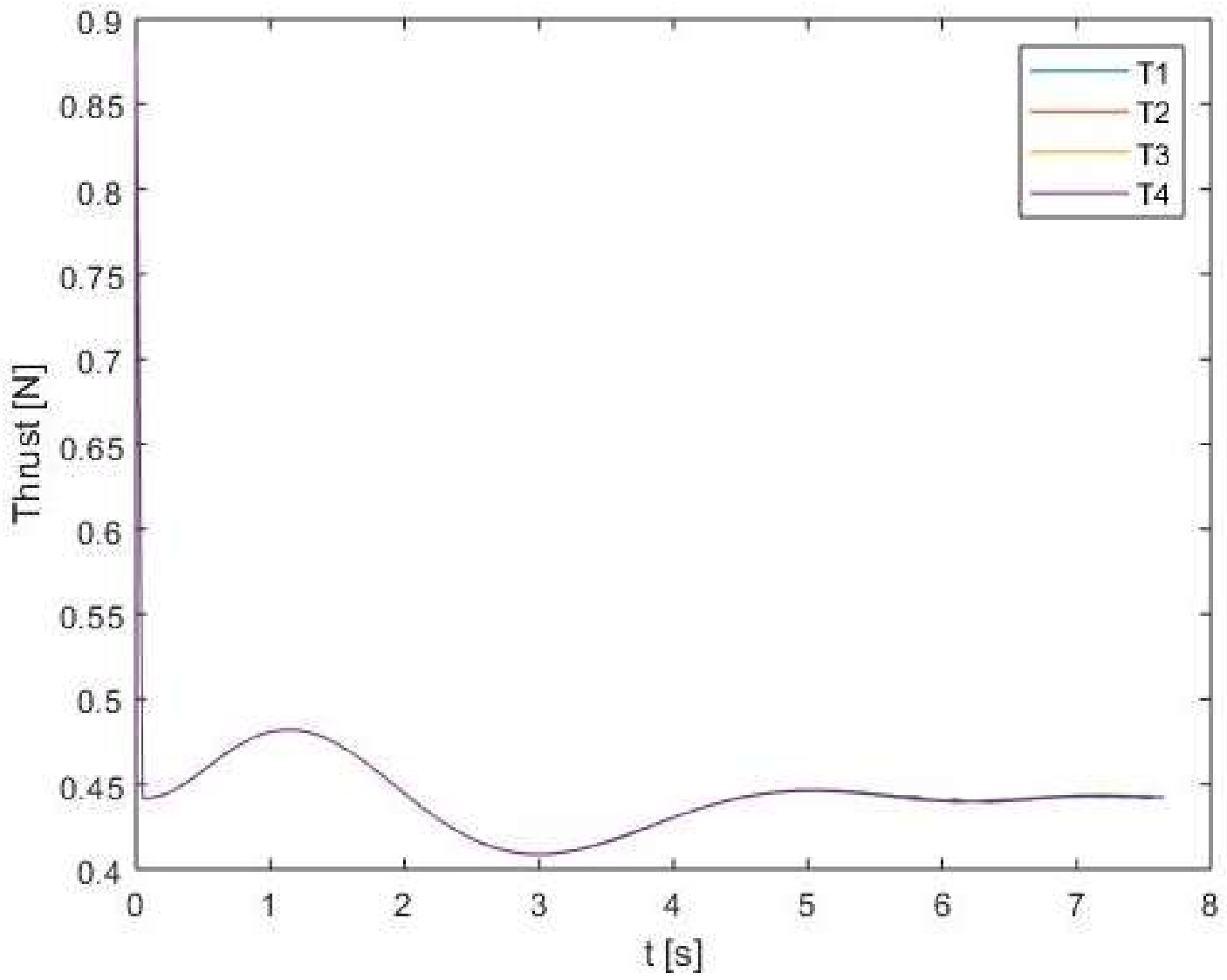

Figure 5.4. The thrusts of the quadrotor

The thrusts are demonstrated in Fig.5.4. There are 4 motors fixed on the quadrotor. And all of them are plotted in the same figure.

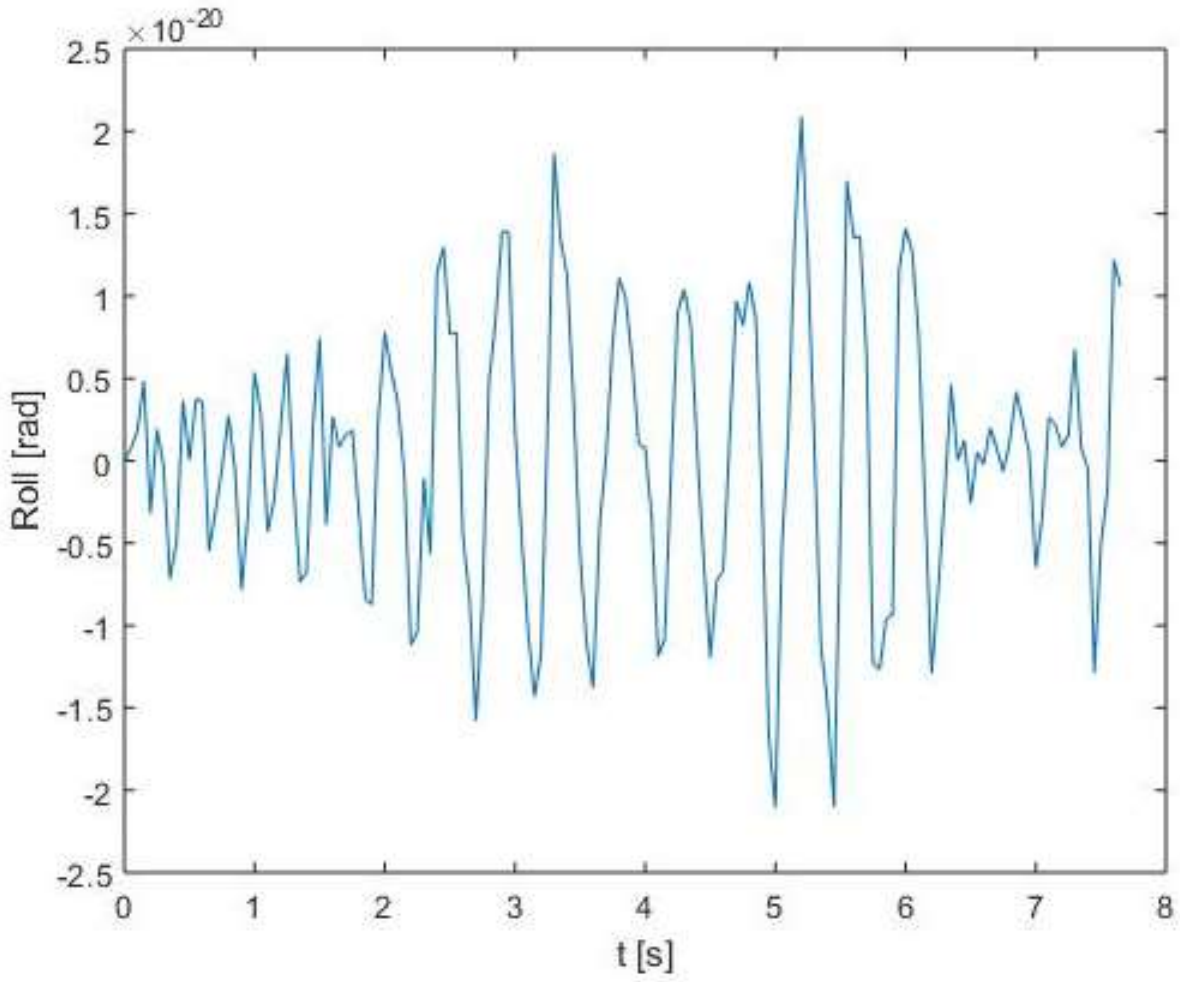

Figure 5.5. Roll of the quadrotor

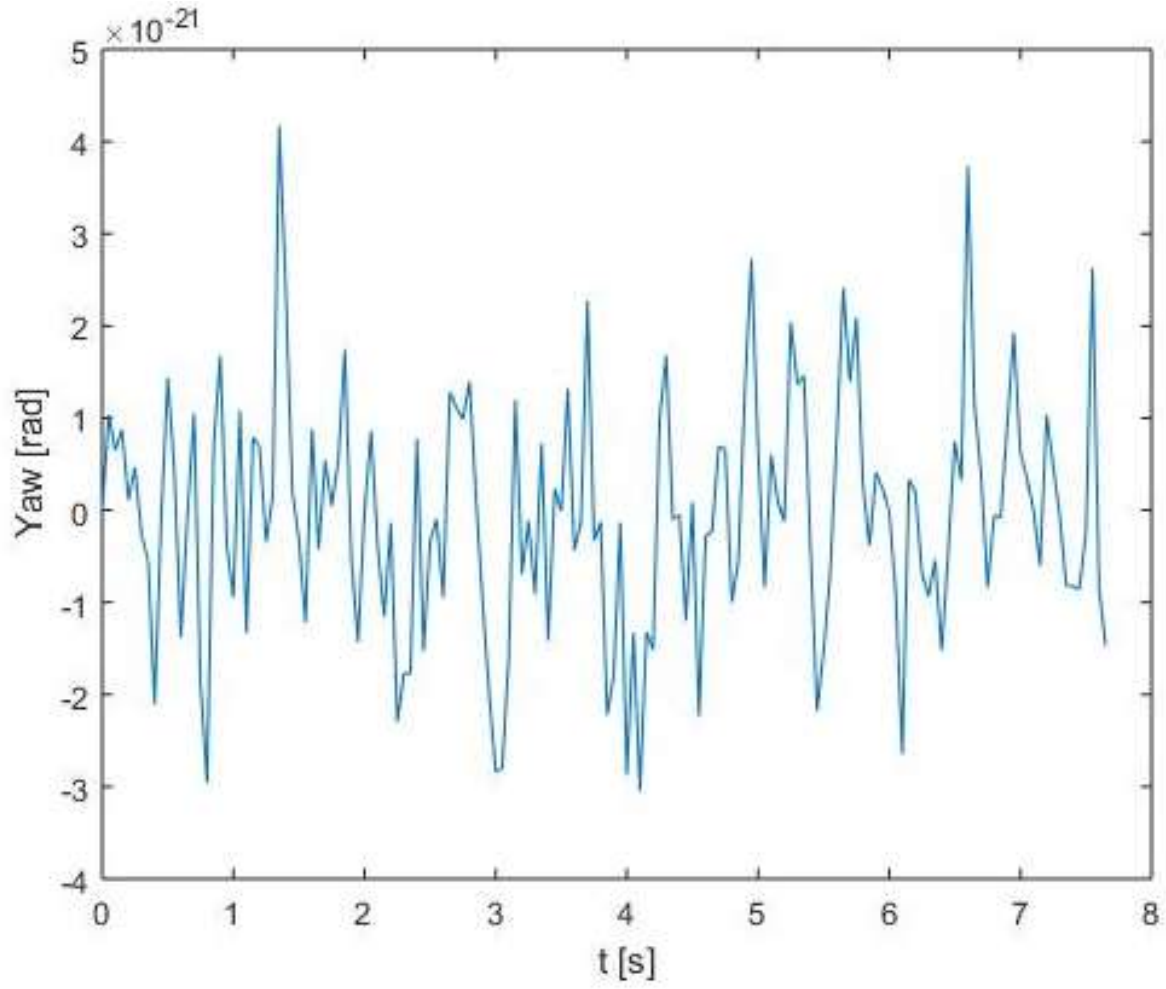

Figure 5.6. Yaw of the quadrotor

The roll and yaw are presented in Fig.5.5. and Fig.5.6. The units of y axis in both figures are 'rad'.

The pitch is recorded in Fig.5.7.

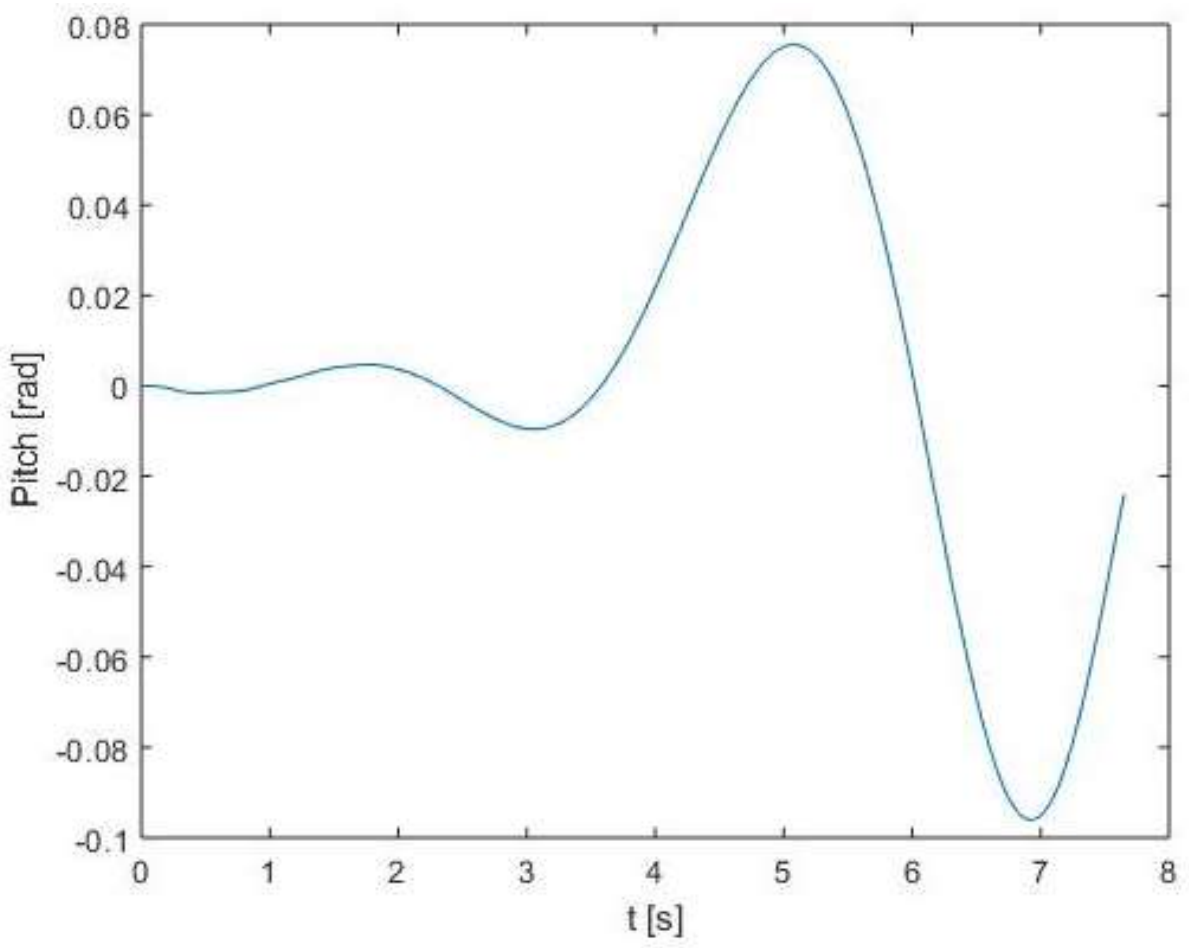

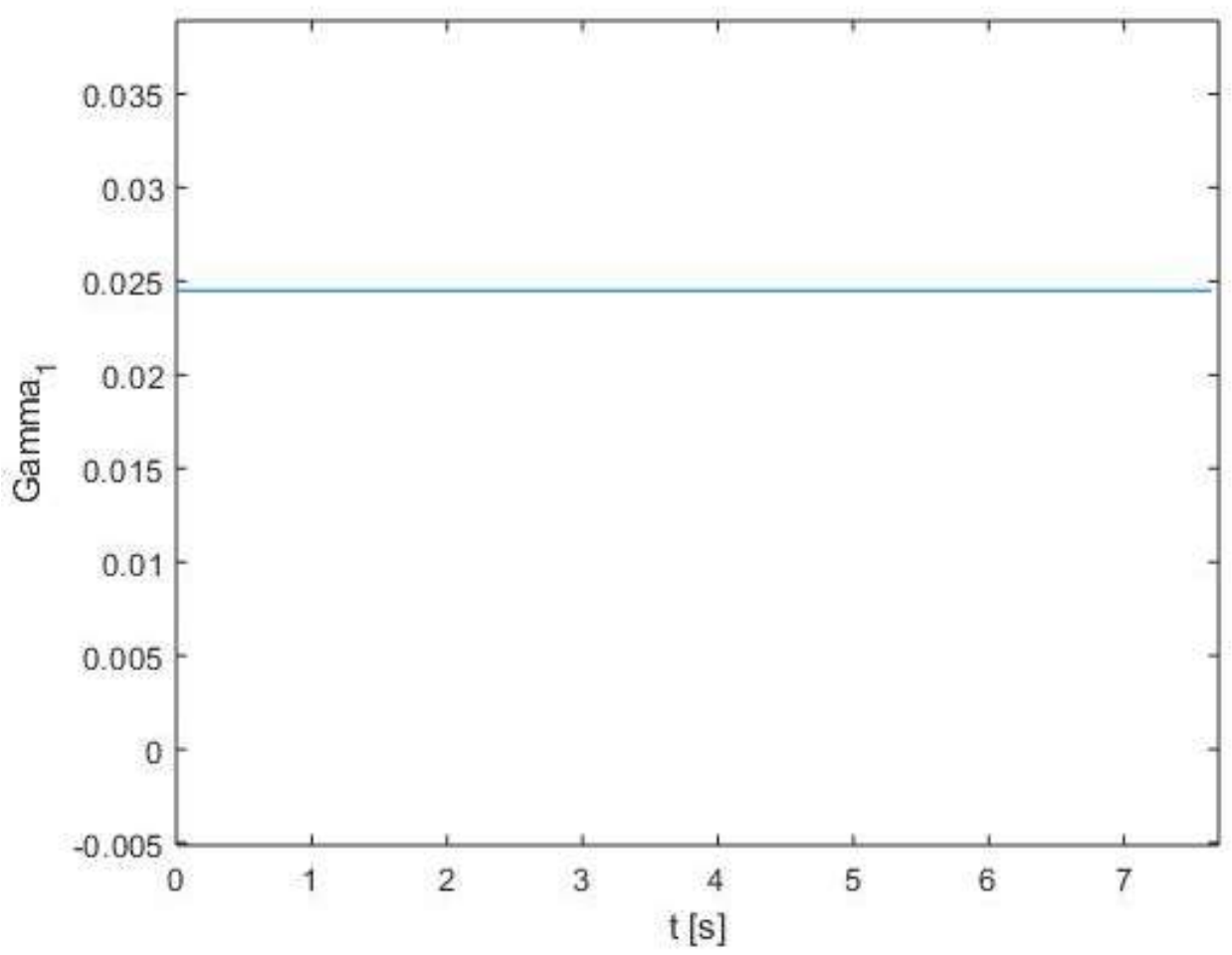

Figure 5.7. Pitch of the quadrotor

Figure 5.8. γ for the first motor

There are four motors. Consequently, there are four γ coefficients in all. As we discussed in (2.8), each of γ is a constant if the temperature is constant. We only plot the γ for the first motor in my thesis. Because the rest 3 coefficients are same in value.

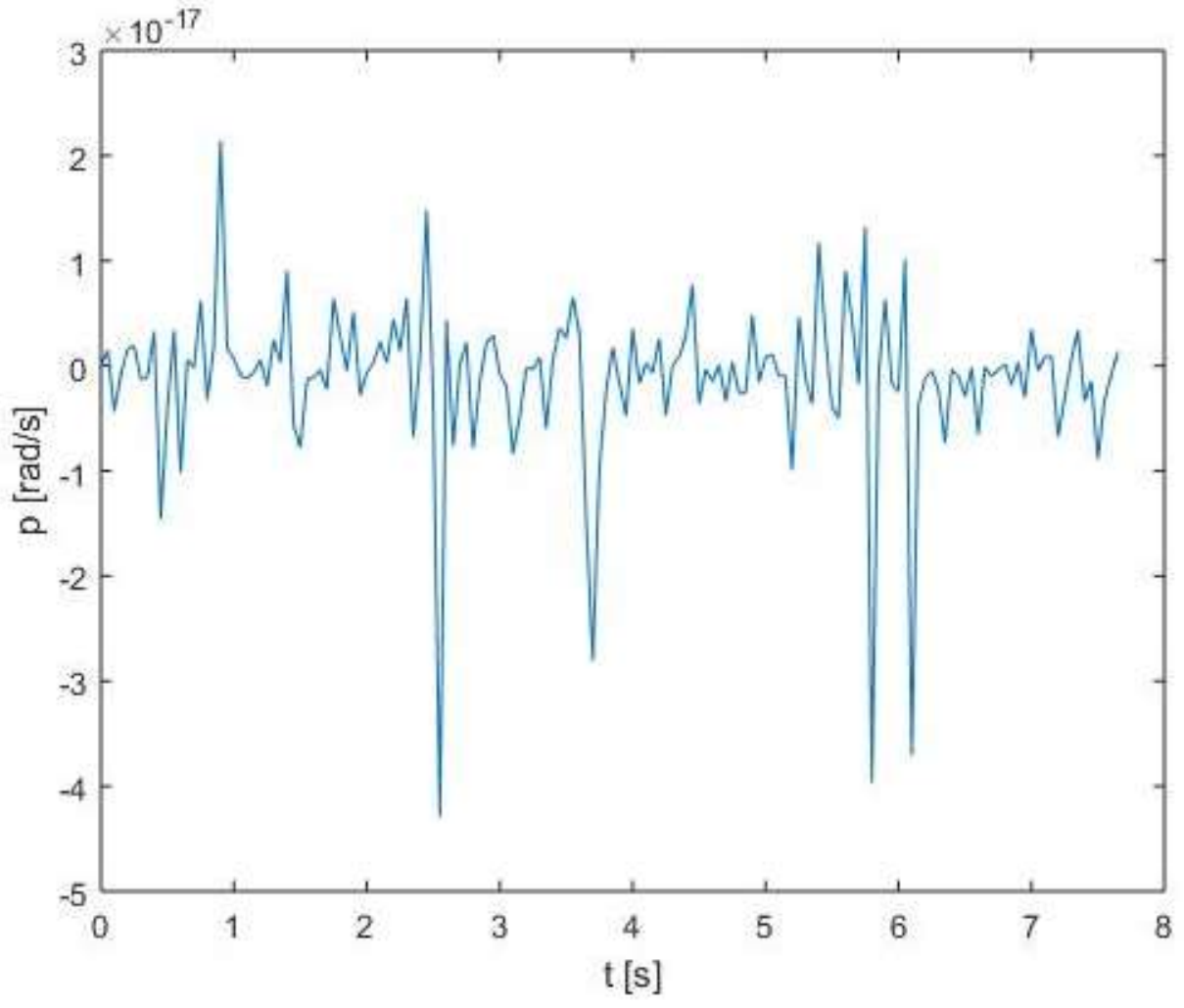

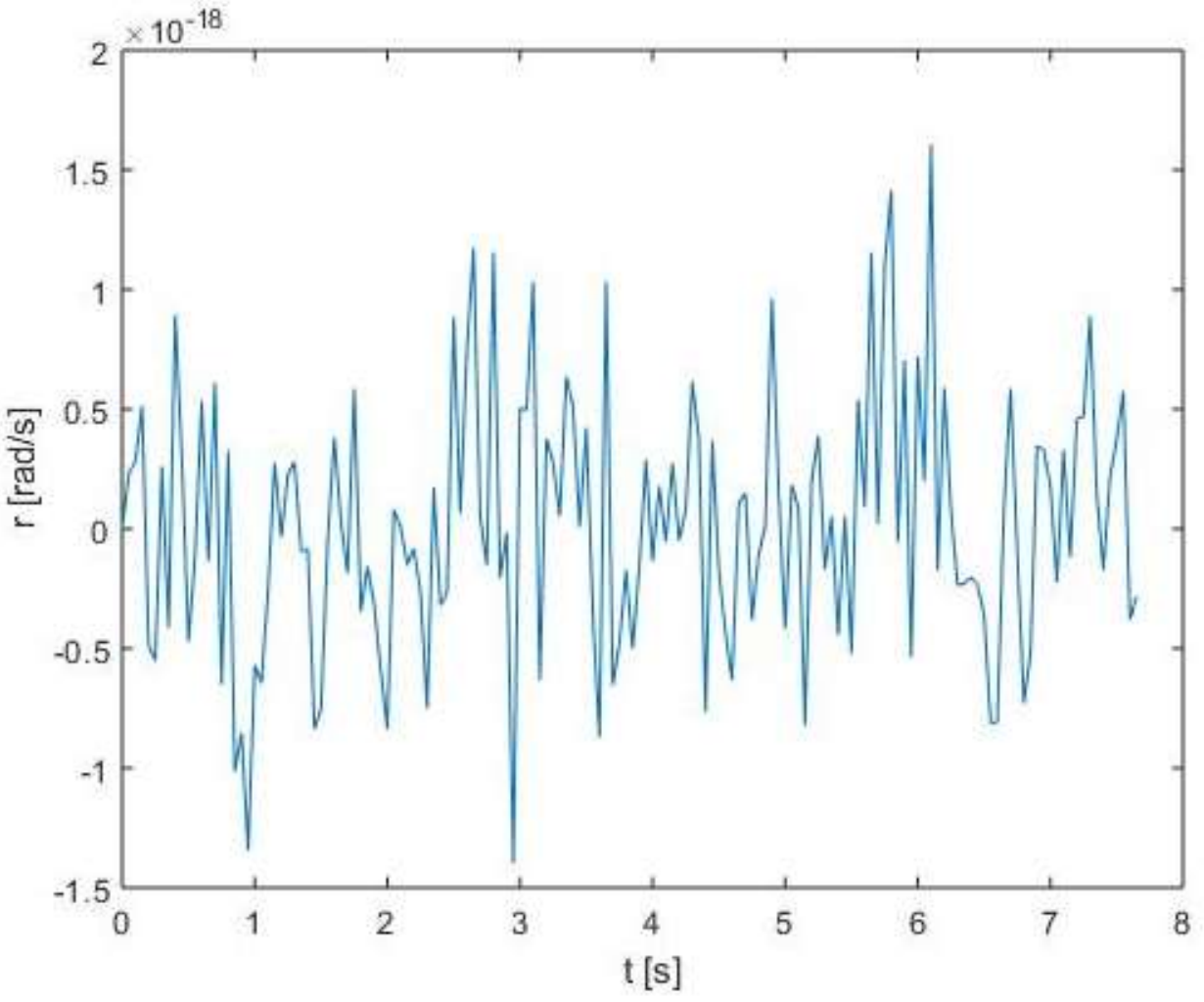

Figure 5.9. Angular velocity along X

Figure 5.10. Angular velocity along Z

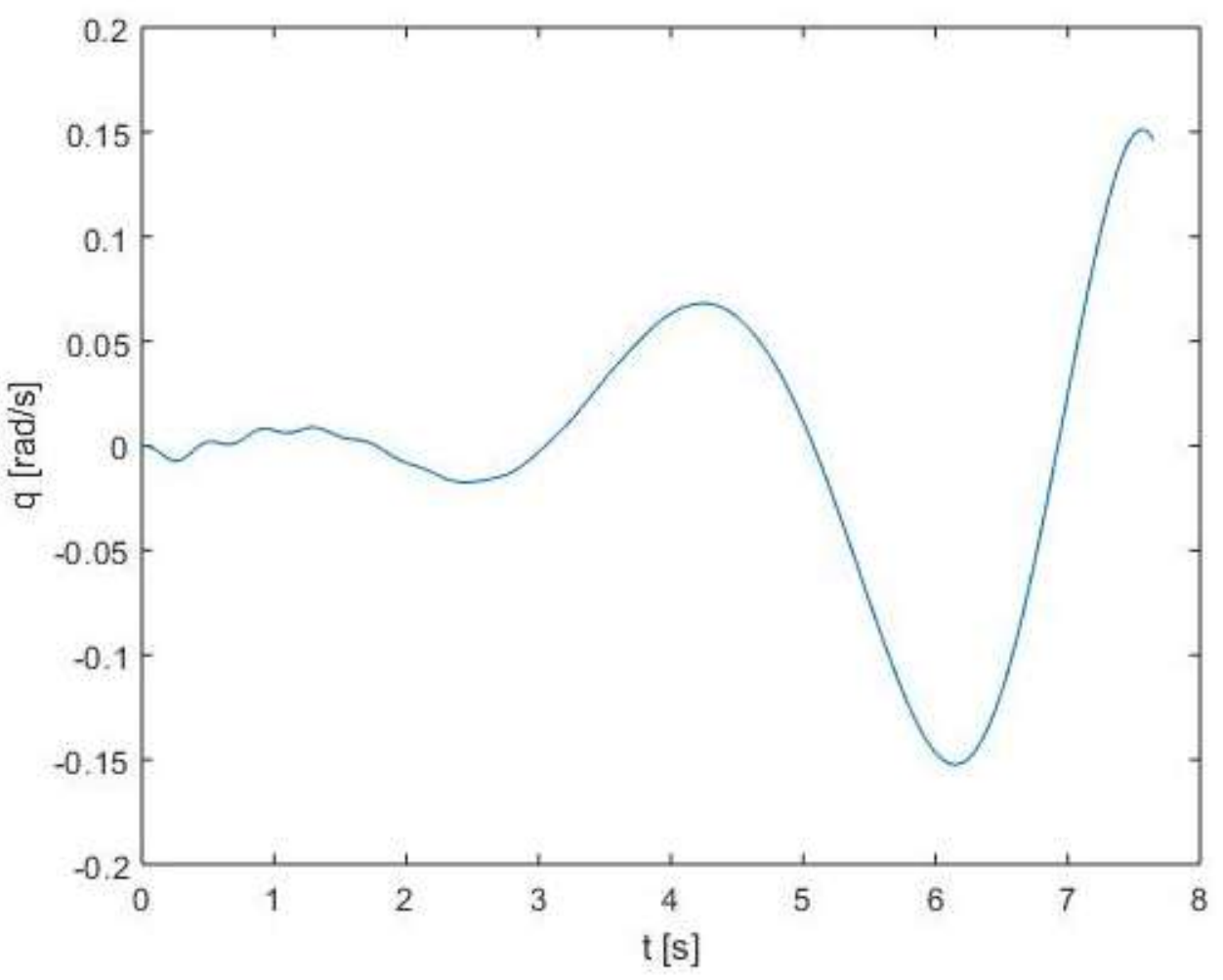

Figure 5.11. Angular velocity along Y

The angular velocities along X-axis, Z-axis, and Y-axis with respective to body-fixed frame are recorded in Fig.5.9. to Fig.5.11.

So far, the simulating results have been fully displayed. And some points of the result are to be elucidated in detail.

The first interesting point is the total moving time. The total time for accomplishing this movement is 7.7 second. While if we look back to path planning, the planned/reference moving time is:

$$\mathrm{t}_{\mathrm{planned}} = \frac{total\ distance}{average\ velocity} = \frac{4}{0.5} = 8.0\ (seconds)$$

The simulator stops at 7.7 seconds however. The reason for this is that the reference generated is very smooth, making the path planned near 8 second changing slightly. The planned position at 7.7s is almost the same to the value at 8.8s; both are 2.00. So, the simulator judges that the quadrotor has already reached the target and stops at 7.7s.

Secondly, let's zoom the z position at 2 second. (Fig.5.12.)

The reference (blue line) is at 1 meter at 2 second. The actual position (red line) of the quadrotor is 0.888 meter. The accomplishment of controlling is 88.8%.

We discussed a lot in 'Altitude Control', Chap 4. We predict mathematically the accomplishment in controlling with a moving reference in this case is 81%.

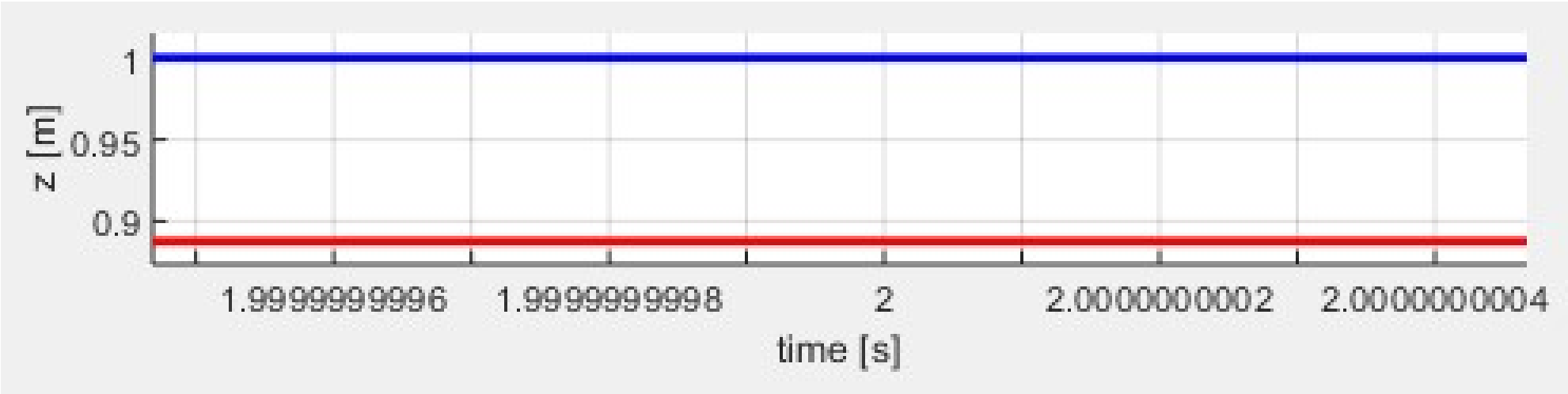

Figure 5.12. Z position at 2 second

The result (88.8%) is quite near to the predict (81%), indicating the correctness in previous work in eigenvalues picking for a moving reference. As mentioned before, the difference of them are contributed by the velocity is not a strict uniform acceleration motion and the system is not strictly first-order.

Additionally, it can be clearly seen that the velocity experienced large bias in Fig.5.3. Because our controller focusses on position only. The velocity is not designed to be followed directly in this research.

Next, the trusts in Fig.5.4 will be discussed. It seems that the value of 4 thrusts are very close, almost making them overlap. We zoom it in Fig.5.13.

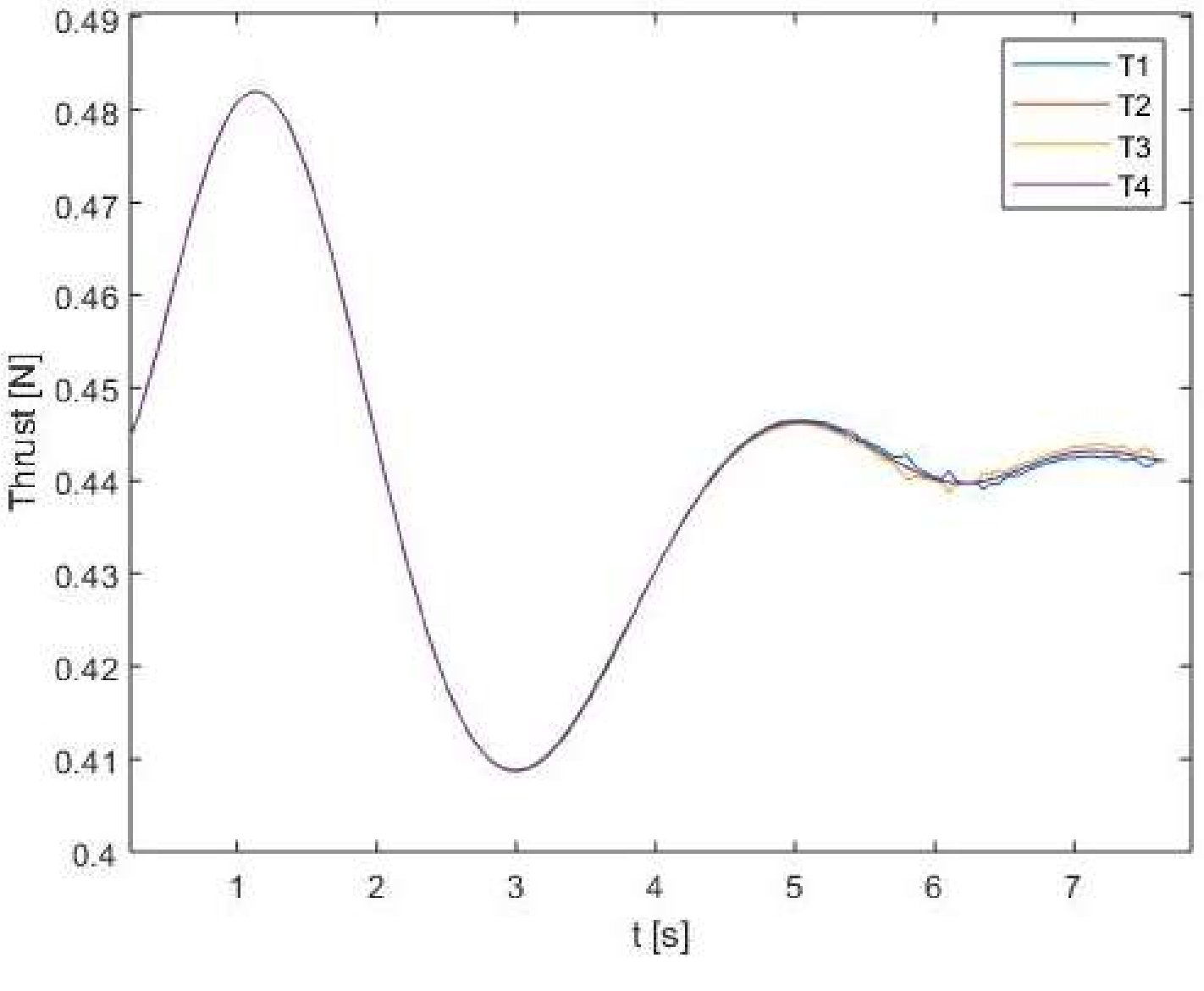

Figure 5.13. Thrusts

It can now be clearly seen that the thrusts are similar before 4 second. After 4s, the thrusts differentiate itself from each other. Another interesting thing is that the thrusts changes in a more intensive way before 4 seconds. While they get milder later after 4s.

The reason is obvious. Before 4 seconds, the quadrotor experiences the process of vertical lifting. Each thrust behaves similar due to symmetry. Also, the thrusts change greatly to generate accelerations to change the height in need. Whereas this change gets mild after 4s when the quadrotor is instructed to translate while maintaining its height. In controlling the attitude, the motors get the different thrust in value.

The translating (moving in a constant height) is achieved by adjusting the pitch of the quadrotor making the thrusts can be decomposed into the desired translating direction. For this reason, the pitch of the quadrotor (Fig.5.7) and angular velocity along Y (Fig.5.11) change greatly after 4s only.

All in all, it is a successful reference tracing controlling. The result yields the way as we expect.

# CHAPTER 6

# Kalman Filter

So far, the quadrotor is well-controlled to some extent where the noise near the volcano is not taken into special consideration. Though the controller should work well with some background noise, it is reasonable if we fuse the noise information into our controller. Especially for the case where the noise can be deduced in detail. With this concern, the Kalman filter is applied to give the best modification of the controller mathematically.

Another reason for utilizing Kalman filter is that the feedback information we used in altitude control in previous chapter are position and velocity of the quadrotor. While the usual sensors provide position and acceleration only. Using Kalman filter can solve this so-called information asymmetry.

In this chapter, the inner noise (sensor noise) and the outer noise (environment noise near the volcano) are to be quantified first. And the Kalman filter for altitude controller is to be addressed after that.

## Noise

The noises are introduced from two sources: the measurement noise from the sensor and the environment noise near the volcano. We will discuss the noise from the sensor first.

The sensors we used in altitude controller are laser scanner and IMU.

Laser scanner measures the height of the quadrotor. The uncertainty in measuring is regarded as measurement noise. The laser scanner we used is Benewake TFmini 0.3-12m 100Hz LiDAR Module Laser Radar Sensor for RC Drone Obstacle Avoidance and Altitude Hold.

The measurement uncertainty is:

$$n_1 = \pm 0.06m \tag{6.1}$$

It is reasonable to model it as a white noise if we divide the bias by 3 and take the result as the standard variance; 99.7% of the data drops in the region of $\pm 3\sigma$.

So, we now model the noise introduced by laser scanner.

$$n_1 \sim N\ (\ 0\ , 0.02^2\ ) \tag{6.2}$$

Also, IMU records current acceleration. We use BNO080 AR VR IMU Nine Axis 9DOF AHRS/IMU Sensor Module. The measurement uncertainty is:

$$n_2 = \pm 0.3m/s^2 \tag{6.3}$$

Similarly, it can be modelled as a white noise:

$$n_2 \sim N\ (\ 0\ , 0.1^2\ ) \tag{6.4}$$

So far, the noises introduced by sensors have been modelled. We are to model the environmental noise next.

As we mentioned in Chapter 1 'The Volcano'. The temperature is changing intensively between 185℃ and 885℃. We will discuss how it influences the dynamics of our quadrotor first.

The thrust is proportional to the air density. While the changing temperature causes the air density inconstant. Consequently, it influences the thrust.

Consider the following relationship (ideal gas state equation):

$$P \cdot M = \rho \cdot R \cdot T \tag{6.5}$$

P is the pressure which is regraded as a constant when the environment is 'freely connected' (The air can flow freely). M is the average Molar mass of the air near the Volcano. It is also a constant. R is Avogadro constant, $6.02 \times 10^{23}$ in value in approximation.

$\rho$ is the density of the air. It is in the unit of $kg/m^3$. T is the temperature of the air. The unit of it is Kalvin. Besides, the relationship between Kalvin and Celsius is:

$$T_{celcius} + 273.15 = T_{Kalvin} \tag{6.6}$$

We have the following relationship:

$$thrust \propto \rho \propto \frac{1}{T} \tag{6.7}$$

or

$$thrust = k_{FT} \cdot \frac{1}{T} \tag{6.8}$$

Reconsider equation (2.1):

$$\ddot{Z} = \frac{1}{m}(F_1 + F_2 + F_3 + F_4) - g$$

This equation matches the situation when temperature is 25℃ $(298.15\text{K})$. And the total thrusts (sum of $F_1$ $to$ $F_4$) can be calculated when the quadrotor is hovering. That is: 1.7658 Newton. The coefficient $k_{FT}$ in (6.8) can be calculated based on this. And the result yields:

$$thrust = 526.4733 \cdot \frac{1}{T} \tag{6.9}$$

While if the temperature increases to 185℃ $(458.15\text{K})$, the corresponding thrust is 1.1491 Newton decreased by 0.6167 Newton compared to the case in 25℃. If the temperature increases to 885℃ $(1158.15\text{K})$, the corresponding thrust is 0.4546 Newton decreased by 1.3112 Newton compared to the case in 25℃.

So that the thrust near the volcano decreases from:

-1.3112 Newton to -0.6167 Newton (6.10)

Based on the fact of changing thrusts, equation (2.1) can be rewritten in the following form:

$$\ddot{Z} = \frac{1}{m}(F_1 + F_2 + F_3 + F_4) - g + n_0 \tag{6.11}$$

Where $n_0$ is to represent the loss of thrust near the volcano. The value is calculated based on (6.10):

$$n_0 \in (-7.2844\,, -3.4261) \tag{6.12}$$

One interesting thing is that the relationship (6.9) transferring temperature to thrust is not linear. So that $n_0$ is not a strict random distribution in (6.12) even if the temperature changing near the volcano follows a strict random distribution.

However, we still deem the distribution of $n_0$ in (6.12) a strict random distribution for further discussion.

Using the similar method, the $n_0$ is modelled as the following white noise:

$$n_0 \sim N\,(-5.3552\,, 0.6430^2\,)$$

-5.3552 is the mean value of the maximum value and minimum value. 0.6430 is calculated by dividing the bias by 3.

So far, the noises have been thoroughly elucidated and modelled.

## Kalman Filter

The Kalman filter is to be designed to fuss the information about these noises.

The equation describing the dynamics and the noise are listed below.

$$\ddot{z} = \frac{1}{m} \cdot (F_1 + F_2 + F_3 + F_4) - g + n_0 \tag{6.13}$$

$$z_m = z + n_1 \tag{6.14}$$

$$\ddot{z_m} = \ddot{z} + n_2 \tag{6.15}$$

Here, (6.13) is same to (6.11).

(6.14) is the equation describing laser scanner. The z is the actual height of the quadrotor. While the measured height $z_m$ by laser scanner contains noise $n_1$ which has been introduced in distribution (6.2).

(6.15) is the equation describing IMU. The $\ddot{z}$ is the real vertical acceleration of the quadrotor. While the measured acceleration $\ddot{z_m}$ by IMU contains noise $n_2$ which has been introduced in distribution (6.4).

Define $n_s$:

$$n_0 = n_s - 5.3552 \tag{6.16}$$

$$n_s \sim N\,(0\,, 0.6430^2\,) \tag{6.17}$$

So that (6.13) can be written as

$$\begin{pmatrix} \dot{z} \\ \ddot{z} \end{pmatrix} = \begin{bmatrix} 0 & 1 \\ 0 & 0 \end{bmatrix} \begin{pmatrix} z \\ \dot{z} \end{pmatrix} + \begin{bmatrix} 0 & 0 & 0 & 0 \\ \frac{1}{m} & \frac{1}{m} & \frac{1}{m} & \frac{1}{m} \end{bmatrix} \begin{pmatrix} F_1 \\ F_2 \\ F_3 \\ F_4 \end{pmatrix} + \begin{pmatrix} 0 \\ -g - 5.3552 \end{pmatrix} + \begin{pmatrix} 0 \\ 1 \end{pmatrix} \cdot n_s$$

(6.18)

Substitute (6.13) and (6.16) into (6.15). The result yields:

$$\ddot{z} = \frac{1}{m} \cdot (F_1 + F_2 + F_3 + F_4) - g - 5.3552 + n_s \tag{6.19}$$

Equations (6.14) and (6.19) can be rewritten in the form of matrix:

$$\begin{pmatrix} \dot{z_m} \\ \ddot{z_m} \end{pmatrix} = \begin{bmatrix} 1 & 0 \\ 0 & 0 \end{bmatrix} \begin{pmatrix} z \\ \dot{z} \end{pmatrix} + \begin{bmatrix} 0 & 0 & 0 & 0 \\ \frac{1}{m} & \frac{1}{m} & \frac{1}{m} & \frac{1}{m} \end{bmatrix} \begin{pmatrix} F_1 \\ F_2 \\ F_3 \\ F_4 \end{pmatrix} + \begin{pmatrix} 0 \\ -g - 5.3552 \end{pmatrix} + \begin{pmatrix} n_1 \\ n_2 + n_s \end{pmatrix}$$

(6.20)

The Kalman coefficient is to be calculated based on (6.18) and (6.20). Additionally, it is worthy of noticing that the noise from the sensor and the noise in the dynamics are corelated to some extent. In other words,

$$E\left( n_s \cdot \begin{pmatrix} n_1 \\ n_2 + n_s \end{pmatrix}^T \right) = (\,0 \quad 0.6430^2\,) \tag{6.21}$$

The function 'lqe' in MATLAB is used to solve Kalman coefficient:

$$k_f = lqe\left( \begin{bmatrix} 0 & 1 \\ 0 & 0 \end{bmatrix}, \begin{bmatrix} 0 \\ 1 \end{bmatrix}, \begin{bmatrix} 1 & 0 \\ 0 & 0 \end{bmatrix}, 0.6430^2, \begin{bmatrix} 0.02^2 & 0 \\ 0 & 0.1^2 + 0.6430^2 \end{bmatrix}, [0 \quad 0.6430^2] \right)$$

(6.22)

The answer yields:

$$k_f = \begin{bmatrix} 3.1434 & 0 \\ 4.9406 & 0.9764 \end{bmatrix} \tag{6.23}$$

So that the Kalman coefficient has been well-designed for the following Kalman filter:

$$\begin{bmatrix} \dot{\hat{z}} \\ \ddot{\hat{z}} \end{bmatrix} = \begin{bmatrix} 0 & 1 \\ 0 & 0 \end{bmatrix} \begin{bmatrix} \hat{z} \\ \dot{\hat{z}} \end{bmatrix} + \begin{bmatrix} 0 & 0 & 0 & 0 \\ \frac{1}{m} & \frac{1}{m} & \frac{1}{m} & \frac{1}{m} \end{bmatrix} \begin{bmatrix} F_1 \\ F_2 \\ F_3 \\ F_4 \end{bmatrix} + \begin{bmatrix} 0 \\ -g - 5.3552 \end{bmatrix} + k_f \cdot \left( \begin{bmatrix} z_m \\ \ddot{z}_m \end{bmatrix} - \begin{bmatrix} \hat{z} \\ \ddot{\hat{z}} \end{bmatrix} \right) \tag{6.24}$$

The height and velocity estimated from Kalman filter in (6.24) are used as the feedback for controlling.

To verify our work, we take this estimator in to our controller. And the MATLAB Simulink is used for verifying.

Assume the quadrotor is supposed to climb to 1 meter from 0 near the volcano, Mt. Satsuma-iojima. The dynamics has been described in (6.18). The reference is fixed at 1 meter. The noises have been introduced in (6.2), (6.4) and (6.17).

Firstly, the controller k is selected by picking a proper eigenvalue pair [-100 -10].

k = place ( A, B, [ -100  -10 ] )

The result yields:

$$k = \begin{bmatrix} 45 & 4.95 \\ 45 & 4.95 \\ 45 & 4.95 \\ 45 & 4.95 \end{bmatrix}$$

The N mentioned in (4.1) can consequently be picked based on (4.7):

$$N = \begin{bmatrix} 45.4415 & 1 \\ 45.4415 & 1 \\ 45.4415 & 1 \\ 45.4415 & 1 \end{bmatrix}$$

Kalman coefficient has been found in (6.23).

The Simulink is demonstrated below: (Fig.6.1)

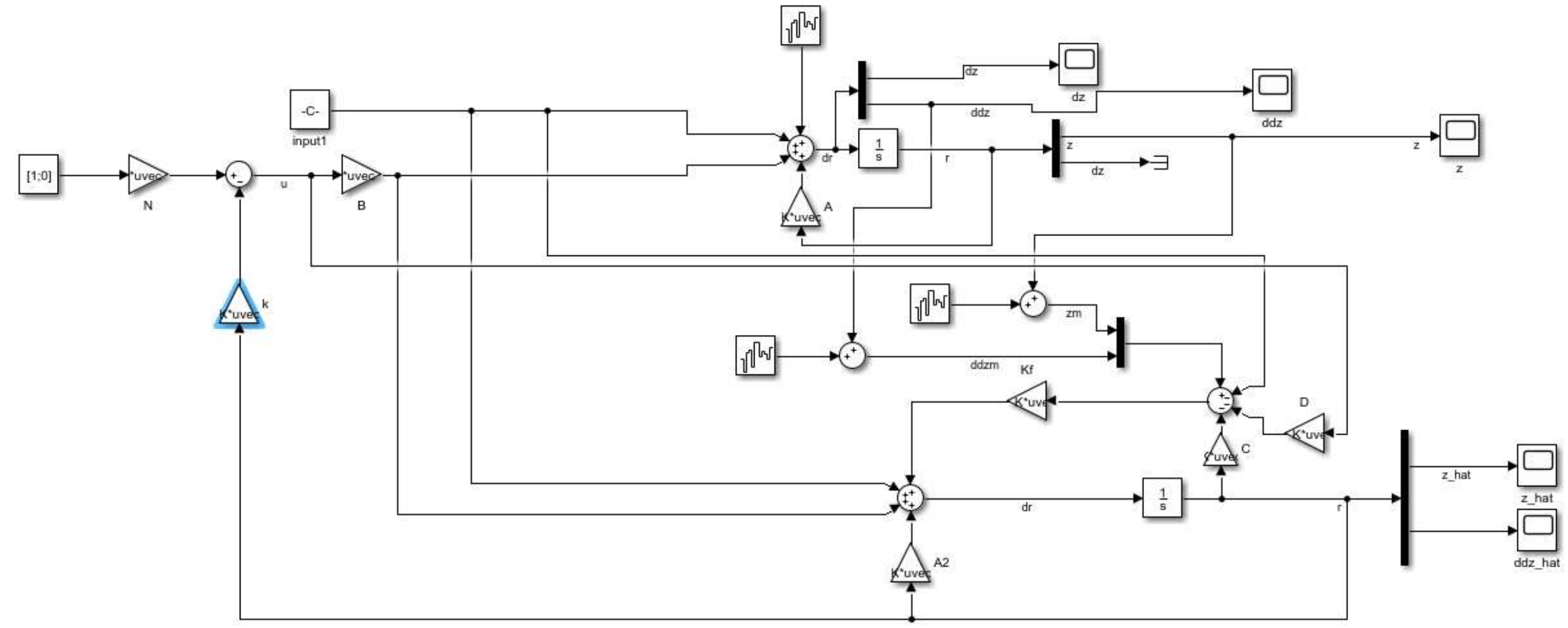

Figure 6.1. Simulink of state space feedback control containing Kalman filter

In general, it consists of 2 sections: the system model as well as the feedback where the feedback is modified using Kalman filter. And the result of it is to be plotted.

The real height of the quadrotor is recorded in Simulink. (Fig.6.2) Note that it is the actual height read from Simulink. We can see the drone goes to 1m as we expect. A large fluctuation occurs in height since noises from dynamics and the sensor have been introduced. Noted that we don't use this data as feedback directly. All the data can only be achieved from sensor in our controlling.

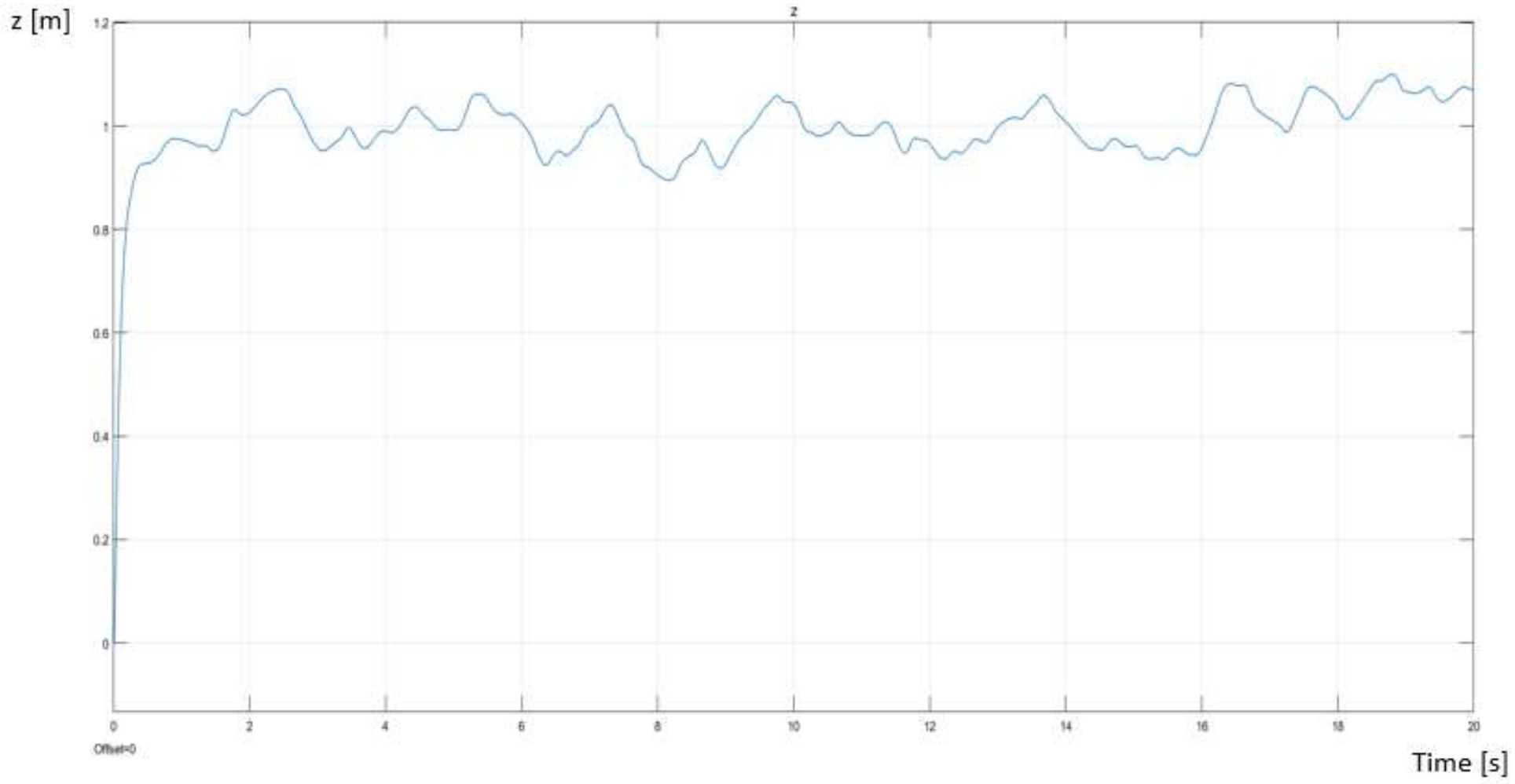

Figure 6.2. Actual altitude of the quadrotor

The readout of the laser scanner recording the height is demonstrated in Fig.6.3.

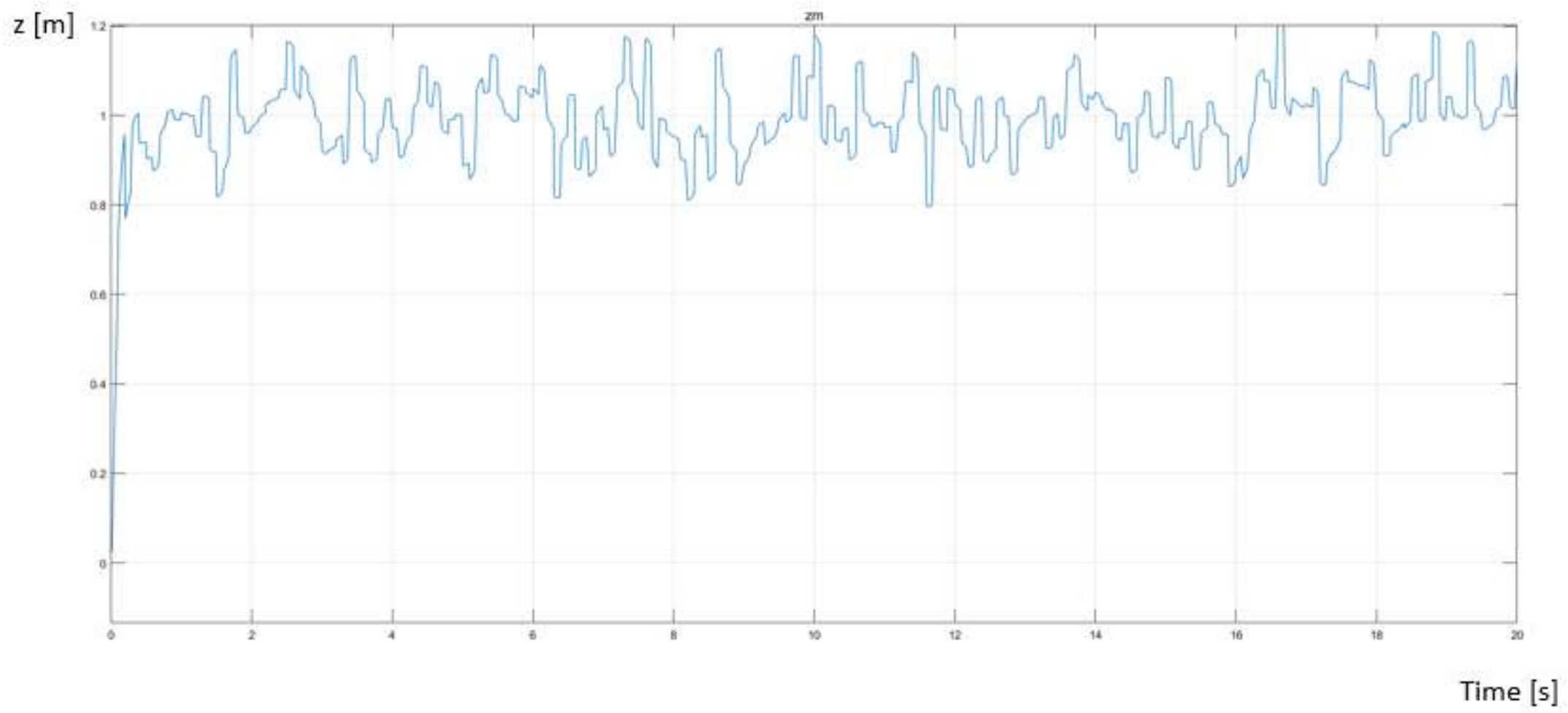

Figure 6.3. Altitude recorded by laser scanner

It can be clearly seen that the fluctuation in the recorded data from sensor is more intensive (higher frequency) compared to the actual height. This is the consequence of addressing measurement noise. It is convincing that we use an observer (Kalman filter) to modify it before using it as feedback.

The height result of Kalman filter has been plotted in Fig.6.4.

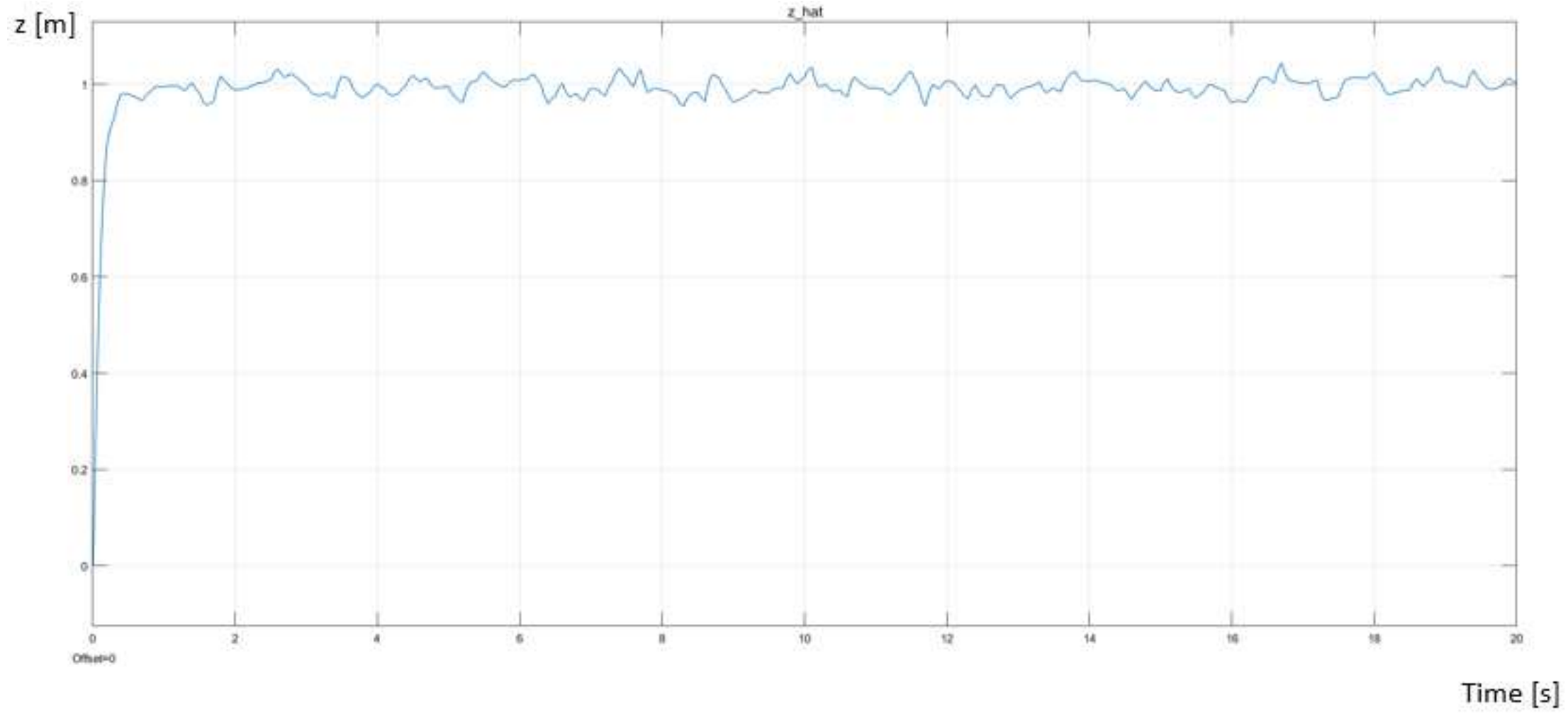

Figure 6.4. Altitude predicted by Kalman filter

The height fluctuates in a less intensive way after applying Kalman filter compared to the readout from sensor. It is understandable since this observer combines the information of noises from sensor and dynamics together.

# Conclusion

As we discussed in the previous chapters, the controller designed is effective in stabilizing the altitude and attitude. In general, the quadrotor goes along the path we designed.

The Kalman filter is effective in combining the information of the changing volcano temperature and the measurement noise. The estimated state works well as the feedback utilized in altitude control.

Moreover, the invented eigenvalue picking method with a moving reference is also effective. The actual accomplishment of movement (88%) is near the estimated value (81%) calculated from eigenvalue picking method.

Further work on eigenvalue picking method with a moving reference is to be modified and published formally and separately after this thesis.